\documentstyle[preprint,aps,amssymb,floats,epsf]{revtex}
\draft
\tighten

\newcommand{\intk}{\int\! {\mkern2mu\mathchar'26\mkern-2mu\mkern-9mud}
{\bf k}\, }

\newcommand{\Z}{{\Bbb Z}}

\newcommand{\f}{\frac}
\newcommand{\half}{\frac{1}{2}}
\renewcommand{\gg}{g_{\rm BCS}}

\newcommand{\Neel}{|\ \hbox{N\' eel}\ \rangle}
\newcommand{\iDelta}{{\it \Delta}}
\newcommand{\iTheta}{{\it \Theta}}

\newcommand{\eq}{\begin{equation}}
\newcommand{\eqend}{\end{equation}}
\newcommand{\eqa}{\begin{eqnarray}}
\newcommand{\neqa}{\begin{eqnarray*}}
\newcommand{\eqaend}{\end{eqnarray}}
\newcommand{\neqaend}{\end{eqnarray*}}
\newcommand{\nonu}{\nonumber \\ \nopagebreak}
\newcommand{\bma}[1]{\begin{array}{#1}}
\newcommand{\ema}{\end{array}}
\newcommand{\bc}{\begin{center}}
\newcommand{\ec}{\end{center}}

\begin{document}

\preprint{Physica C {\bf 296}, 119-136 (1998).}

\title{Mean field analysis of a model for superconductivity in an
antiferromagnetic background}

\author{Edwin Langmann$^a$, Jack Lidmar$^a$, Manfred Salmhofer$^b$,
and Mats Wallin$^a$}
\address{$^a$ Theoretical Physics, Royal Institute of Technology,
S-100 44 Stockholm, Sweden \\ $^b$ Mathematik, ETH Zentrum, 8092
Z{\"u}rich, Switzerland\\
\date{\today}
}

\maketitle
\begin{abstract}
We study a lattice fermion model for superconductivity in the presence
of an antiferromagnetic background, described as a fixed external
staggered magnetic field.  We discuss the possibility that our model
provides an effective description of coexistence of antiferromagnetic
correlations and superconductivity, and its possible application to
high temperature superconductivity.  We also argue that, under certain
conditions, this model describes a variant of the periodic Anderson
model for heavy fermions.  Using a path integral formulation we
construct mean field equations, which we study in some detail.  We
evaluate the superconducting critical temperature and show that it is
strongly enhanced by antiferromagnetic order.  We also evaluate the
superconducting gap, the superconducting density of states, and the
tunneling conductivity, and show that the most stable channel
usually has a $d_{x^2-y^2}$-wave gap.
\end{abstract}


\section{Introduction}

Motivated by heavy--\-fermion-- and high--\-tempe\-ra\-ture
super\-con\-ductors, considerable theoretical effort has gone into
explaining the occurrence of unconventional superconductivity in
strongly correlated electron systems.  A basic model of interest here
is the one-band Hubbard model which describes lattice fermions
interacting with a strong on--site Coulomb repulsion \cite{Dagotto}.
This model is a widely studied candidate for explaining the unusual
properties of the high temperature superconductors.  Despite massive
efforts this model is not yet well understood due to the technical
difficulties involved.  It is thus desirable to devise approximate
simplified models focusing on certain aspects of the physics, that are
easier to analyze than the full model, and whose properties can be
compared with experiments and simulations.  In this paper we study
such a model focusing on the possible influence of antiferromagnetism
on superconductivity.

There are several reasons for studying the coupling between
antiferromagnetic (AF) correlations and superconductivity (SC):
(i) The half filled Hubbard model is known to be an insulating
antiferromagnet, and quantum Monte Carlo results \cite{Dagotto} and
results from Hartree-Fock theory \cite{HartreeFock} indicate that
local AF correlations persist up to quite high doping levels (at least
for parameters adequate for high temperature superconductivity).
(ii) The parent compounds of the high-temperature superconductors
(HTSC) are Mott insulators with antiferromagnetic long range order.
Neutron scattering experiments~\cite{Neutron} show that the
antiferromagnetic correlation length gets reduced upon doping.  NMR
measurements, however, suggest that strong short range AF correlations
remain important all the way into the SC phase (see Ref.~\cite{Zha}
for a recent discussion on this point).
(iii) As we will further discuss below, AF correlations may also play
a role in the periodic Anderson model for heavy fermions
\cite{Hewson}.
(iv) Finally there are other (low--$T_c$) superconductors which
exhibit simultaneous SC and AF order \cite{Gig,Maple}.

Several different approaches to consider the effects of AF
correlations on HTSC have been explored in the literature.  One
frequently studied model is based on fermion pairing due to exchange
of AF spin fluctuations \cite{spinfluc}.  Another is the spin bag
mechanism which is based on the idea that local suppression of the AF
gap leads to an effective attraction between certain quasiparticles,
and thus to superconductivity \cite{spinbag}.  Recently Dagotto and
coworkers proposed and studied weak coupling models of holes in an
antiferromagnet, using a numerically determined dispersion relation
\cite{Dagotto1}.  Most of these models lead to $d_{x^2-y^2}$ pairing,
which indeed seems to be observed in experiments on HTSC \cite{dwave}.

The model studied in this paper describes lattice fermions in an
external antiferromagnetic field coupled to the fermion spin.  We also
include various attractive, instantaneous fermion-fermion couplings
(see Sec.\ \ref{model1}).  Although we use a field with AF long-range
order, we argue below that, for the properties we calculate, this
model gives a good approximation also when the AF order parameter is
slowly varying in the sense further discussed below.  Our model
illustrates in a simple way how correlations can narrow the
quasiparticle bands which increases the density of states (DOS) and
hence increases $T_c$ \cite{KM,Machida}.  This is an example of the
van Hove scenario which has been frequently discussed in the context
of HTSC \cite{vH,Dagotto1}.  In our model the enhancement of the van Hove
singularity by antiferromagnetism increases $T_c$, such that the right
order of magnitude for HTSC can be obtained by reasonably small
couplings.  Recent numerical results suggest that in the parameter
range of interest for HTSC, the Hubbard model has a DOS peak, and the
chemical potential intersects this peak for certain finite doping
\cite{Bulut}.  While it is clear that the simple band structure
obtained in our model is not going to be realistic in detail, our
model attempts to (i) represent the states at this peak by the AF
bands, and (ii) to model the resulting superconducting properties.
Since our model is simple, our investigation can be done in some
detail which would be difficult for more realistic models. Our
approach is similar to Ref.\ \cite{Dagotto1} in that we attempt to
model quasiparticle bands renormalized by strong correlations and then
investigate their superconducting properties, but otherwise the models
are quite different.

The model we study naturally leads to a superconducting gap $\Delta$
with a staggered real space structure (i.e., with a two component
order parameter).  A superconducting order parameter with such a
structure has been studied previously on a phenomenological level in
the context of heavy fermions~\cite{Heid}.  Our results show how such
an order parameter can arise in microscopic models.

The first purpose of this paper is to provide motivation for our model
and develop efficient tools to study it.  In any case, our model
should describe systems where antiferromagnetic order and
superconductivity coexist \cite{Maple}.  However, we also argue that
it can be regarded as a useful toy model for studying the effects of
band renormalization by strong correlations on superconductivity in
HTSC or heavy fermion systems.  The idea is that by using a
Hubbard--Stratonovich transformation, an interacting fermion systems
can be represented as non--interacting fermions coupled to a {\em
dynamical} boson field.  A simple approximation is to replace the
dynamical boson field by a fixed boson field, which is determined by
Hartree-Fock (HF) equations.  A general feature of the doped Hubbard
model is that this HF field has a nontrivial spatial dependence
\cite{HartreeFock}.  This leads to a variety of complicated bands.
The model studied in this paper represents the simplest nontrivial
case where this occurs.  To study this model, we use an efficient path
integral formalism which should be useful also in more complicated
cases, and we derive mean field equations of BCS type.  To further
investigate the relevance of the description provided by our model a
more elaborate analysis (e.g.\ fluctuation calculations) is required,
but this is beyond the scope of this paper.

The second purpose of the paper is to calculate some physical
properties of the model.  Using parameters motivated by HTSC, we
present a systematic analysis of the different possible SC channels.
We study the effect of on--site, nearest neighbor (nn) and
next--nearest neighbor (nnn) interactions.  We find that the
dominating channels are neither translation-- nor spin rotation
invariant.  In the presence of AF correlations, we find that on-site
attraction never leads to significant pairing and nn attraction is
most efficient and usually produces a $d_{x^2-y^2}$-wave gap,
$\iDelta\propto\cos(k_1) -\cos(k_2)$.  For dominant nnn-attraction we
get $p$-wave pairing, $\iDelta\propto
\sin(k_1+k_2)+ \sin(k_1-k_2)$.  $d_{x^2-y^2}$-wave superconductivity
has been found in numerous other calculations \cite{dwave}, but our
result shows that it arises already in a particularly simple model.
We also derive and solve the gap equations for nn coupling in the full
temperature range below $T_c$.  Our formalism is convenient to
evaluate SC properties.  As an example, we evaluate the SC density of
states and the resulting tunneling conductance as a function of
temperature.  We note that some features of our results appear
qualitatively similar to HTSC, e.g.\ the doping dependence of $T_c$
and the tunneling conductance.

The plan of this paper is as follows.  In Section~\ref{model} we give
a detailed description of the model, and then discuss the possibility
that it may provide a simple description of certain aspects of the
physics of the Hubbard model with an additional weak pairing
interaction.  We also relate our model to the periodic Anderson model
for heavy fermions \cite{Hewson} Section \ref{pa}.  Our analysis of the
model starts in Section~\ref{formalism}, where we summarize the path
integral formalism.  The mean field equations are obtained in
Section~\ref{MFeqs}.  We first (Sec.~\ref{Tceqs}) obtain the
$T_c$-equations for all possible superconducting channels (singlet,
and triplet, translational-invariant and staggered, $s$-, $p$- and
$d$-wave) which are needed in our systematic stability analysis, and
then (Sec.~\ref{btc}) derive the gap equations below $T_c$ for the
dominating channels.  Section~\ref{numres} contains our numerical
results of our stability analysis, the temperature dependence of the
SC gap, and the density of states and the tunneling conductance as a
function of $T<T_c$.  We end with conclusions in
Section~\ref{conclusion}.  Some technical details are in two
appendices.

\section{The model}\label{model}

In this section we present our model (A).  We then discuss the relation of
our model to the doped Hubbard model (B) and the periodic Anderson model
(C).

\subsection{Model description}\label{model1}

Our model Hamiltonian describes lattice fermions with field operators
$a^{(+)}_{\uparrow,\downarrow}({\bf x})$ in an external staggered magnetic
field ${\bf B}({\bf x})$ and an additional weak attractive BCS-like
interaction,
\eq
H_{\rm eff}=H_{\rm hop} +\sum_{{\bf x}} {\bf S}({\bf x})\cdot {\bf B}({\bf
x})\, e^{{i}{\bf Q}\cdot{\bf x}}  +H_{\rm int}
\label{mdl}
\eqend
where ${\bf B}=(B_1,B_2,B_3)$ is the staggered magnetic field
representing the AF order,
\eq
H_{\rm hop}= - \sum_{\langle{\bf x},{\bf y}\rangle, \sigma} t\,
a^+_\sigma({\bf x})a_\sigma({\bf y})
\eqend
is the usual hopping Hamiltonian (i.e.\ the sum over all
nearest-neighboring sites $\langle{\bf x},{\bf y}\rangle$),
${\bf S}=a^+({\bf x}){\mbox{\boldmath $\sigma$} } a({\bf x})$
is 2 times the electron spin operator ($\mbox{\boldmath $\sigma$}$ are
the Pauli spin matrices), and
\begin{mathletters}
\eq
\label{VBCS}
H_{\rm int} = \f{1}{2} \sum_{{\bf x},{\bf y}} n({\bf x}) V({\bf x}-{\bf y})
n({\bf y}) ,
\quad n=\sum_\sigma a_\sigma^+a_\sigma
\eqend
is a (weak) attractive charge-charge interaction.  We consider the model
on a $d$-dimensional cubic lattice $\Z^d$ (i.e.\ spatial vectors are
${\bf x}=(x_1,\ldots, x_d)$ with $x_i$ integers), and  ${\bf Q} =
(\pi,\ldots,\pi)$ is the usual AF vector.
We use a pairing potential with the Fourier transform
\eq
\label{pot}
V({\bf k}) = -g_0 - g_{\rm nn}\sum_i 2\cos(k_i) -
g_{\rm nnn}\sum_{i\neq j} 2\cos(k_i)\cos(k_j)
\eqend
\end{mathletters}
where ${\bf k}=(k_1,\ldots , k_d)$ with $-\pi\leq k_i\leq\pi$ and
$g_i>0$.  The first, second and third terms here describe on-site,
nearest neighbor (nn) and next-to-nearest neighbor (nnn) instantaneous
interactions.  We take the corresponding couplings $g_i$ as input
parameters of our model.  A useful way to think about this potential
is as coming from a Taylor expansion of some general potential $V({\bf
x}-{\bf y})$, where only the lowest order terms are kept since they
are the only ones expected to be important for SC.  Note that the
normalization in Eq.\ (\ref{pot}) is such that $V({\bf x}-{\bf y})$ is
$-g_0$, $-g_{\rm nn}$ and $-g_{\rm nnn}$ if ${\bf x}$ and ${\bf y}$ are equal,
nn, and nnn, respectively.  We do not fix dimension in our formal
manipulations, but in our numerical calculations we take $d=2$.

We will analyze the model setting ${\bf B}({\bf x})$ independent of
${\bf x}$.  This case is of interest for materials where AF long range
order coexisting with SC has been observed \cite{Maple,Gig}.  However,
we expect that our results are also adequate for the case of varying
staggered magnetic fields ${\bf B}({\bf x})$, as long as this
variation is slow.  Arguments for this will be given below.  In
particular, even though the second term in Eq.\ (\ref{mdl}) explicitly
breaks spin rotation invariance, this model can nevertheless describe
also situations where this symmetry is unbroken:
one can think of Eq.\ (\ref{mdl}) as coming from an adiabatic
approximation (similar to the Born--Oppenheimer approximation) of a
strong--coupling model in which the dynamical AF fluctuations are
frozen, and spin rotation invariance will be restored after averaging
over certain ${\bf B}({\bf x})$--configurations.

\subsection{On the relation to the doped Hubbard model}
\label{htsc}

Here we describe how our model can be related to the doped Hubbard
model with an additional attractive interaction $H_{\rm int}$ as in
Eq.\ (\ref{VBCS}).  The idea is to replace the strong Coulomb
interaction by coupling the fermions to dynamical bosons.  In the path
integral formalism this can be done by a Hubbard-Stratonovich
transformation which introduces dynamical boson fields \cite{HS} \eq
\label{Neel} \phi_0(\tau,{\bf x}) = r(\tau,{\bf x}),\quad
\mbox{\boldmath $\phi$}(\tau,{\bf x}) =
{\bf B}(\tau,{\bf x}) \, e^{{i}{\bf Q}\cdot{\bf x}}
\eqend
depending on imaginary time $\tau$. These fields have a simple
physical interpretation: $s=|{\bf B}|$ and ${\bf e} = {\bf B}/|{\bf
B}|$ represent the magnitude and direction of the AF ordered fermion
spins and $r$ the fermion charge.

A standard approximation then is to replace this dynamical boson
fields by a fixed and static boson field configuration.  This formally
amounts to a saddle point evaluation of the exact boson path integral.
The stationary points in this path integral are determined by HF
equations, and this procedure corresponds to mean field theory.  For
the doped Hubbard model this approximation is already quite involved:
Numerical results show that the HF solutions have a complicated
spatial dependence and, depending on parameters, describe magnetic
domain walls, magnetic vortices etc.\ \cite{HartreeFock}.  However, in
the parameter regime of interest all these solutions exhibit
antiferromagnetic correlations.  The non--trivial assumptions required
to establish a connection of our model to the Hubbard model are (i) HF
theory is adequate to determine interacting fermions bands, (ii) it is
mainly these antiferromagnetic correlations which determine these
bands close to the Fermi surface.

Let us be more specific on point (ii): For thin domain walls or
localized vortices (e.g.), $r({\bf x})= r$ and ${\bf B}({\bf x})={\bf
B}$ are constant in large regions of configuration space.  Then there
are fermion states which, for most ${\bf x}$, coincide with states
describing a system with constant $r({\bf x})= r$ and ${\bf B}({\bf
x})={\bf B}$.  Theses states can be viewed as scattering solutions,
e.g., in a domain wall background.  Our model tries to capture the
effect of AF correlations on such delocalized states.  Our assumption
is that such states exist and that for suitable doping, they are at
the Fermi energy.  Note that the filling $\rho$ used in our model
should not be interpreted as filling in the Hubbard model: $\rho-1$ is
the doping of the AF band, which is smaller than the total doping if
part of the charge carriers are bound, e.g., in domain walls.  A check
of the properties we assumed requires a rather detailed analysis of
the Hubbard model itself and is beyond the scope of this paper.  These
properties seem however to have some support from recent numerical
simulation results for the Hubbard model, which indicate that for
parameters adequate for HTSC, $\mu$ indeed intersects a peak in the
DOS at a certain doping level \cite{Bulut}.

We analyze the model setting ${\bf B}(x)$ independent of $x=(\tau,{\bf
x})$, but we expect that our results are a good approximation also if
${\bf B}(x)$ varies slowly or only in small fractions of
spacetime. Then the 2-point Green functions of our model can be
approximated by
\eq
G(x,y) \approx G_N(x-y,y) + G_A(x-y,y)\, e^{{i}{\bf Q}\cdot{\bf y}}
\eqend
where $G_{N,A}$ depends on the second argument $y$ only via ${\bf
e}(y)\cdot\mbox{\boldmath $\sigma$}$ and $s(y)$.  These Green
functions can be obtained from the ones for ${\bf B}(y)=s\, {\bf
e}=const$ in leading order of a gradient expansion.  In particular,
thermodynamic properties only depend on $s(y)$ in this approximation.
If $s(y)\approx const$, we thus expect that the thermodynamic
properties of our model should not depend much on the AF correlation
length (assuming the latter is large enough).  If $s(y)$ also varies
in small fractions of spacetime, the thermodynamic properties are
appropriate averages of the `local' properties which can be deduced
from the results in this paper.

\subsection{On the relation to the periodic Anderson model}
\label{pa}

Here we present an argument that our model, with constant ${\bf B}$,
can be obtained from the periodic Anderson model in a certain
parameter regime.  The physical picture is as follows: the periodic
Anderson model is a 3D lattice fermion model with two kinds of
spin-$\f{1}{2}$ fermions: $f$-electrons with a strong Hubbard
repulsion are coupled weakly to non-interacting $a$-electrons
\cite{Hewson}.  For large on-site repulsion, the $f$-electrons can be
regarded as half filled and antiferromagnetically ordered.  The lower
magnetic $f$-band is then far underneath the Fermi surface and the
upper one far above.  Thus the $f$-electrons are in a half--filled
band and thus act as a commensurate [${\bf e}(x) = const$] AF
background for the weakly coupled $a$-electrons, and our model
provides an effective weak-coupling description of these
$a$-electrons.

We now turn to a more detailed argument.  We start from the following
periodic Anderson Hamiltonian
\begin{mathletters}
\eq
H=H_{\rm hop}+H_f+V_{fa}+ H_{\rm int}
\eqend
where $H_{\rm hop}$ and $H_{\rm int}$ are as above, and
\eq
H_f= \sum_{{\bf x}} \Bigl(-\epsilon_f n_f({\bf x}) +
U \, n_{f,\uparrow}({\bf x})n_{f,\downarrow}({\bf x}) \Bigr)
\eqend
describes another species of electrons denoted as $f$ which are localized
and have a strong on-site charge-charge repulsion $U>>\epsilon_f>>t$
($n_{f,\sigma}({\bf x}) = f_\sigma^+({\bf x})f_\sigma({\bf x})$,
and $n_f = n_{f,\uparrow}+n_{f,\downarrow}$).
The interaction of the $a$ and $f$ electrons is local,
\eq
V_{fa} = \lambda\sum_{{\bf x}} \Bigl( a^+_\sigma({\bf x})f_\sigma({\bf x}) +
f^+_\sigma({\bf x})a_\sigma({\bf x})\Bigr)\, ,
\eqend
\end{mathletters}
and $H_{\rm int}$ is as in Eq.\ (\ref{VBCS}) i.e., we assume the BCS
attraction among the $a$ electrons only.  $H_a+H_f+V_{fa}$ is the
Anderson Hamiltonian: $H_f$ is the usual strongly repulsive Hubbard
Hamiltonian (with hopping constant $t_f\to 0$) which in the
half-filled case (one $f$-electron per site) describes
antiferromagnetically ordered electrons.  The $a$-electrons are
assumed in the conduction band weakly coupled to the $f$-electrons,
and since $U>>\epsilon_f>>\lambda^2$ and $\epsilon_f >> t$, one still
expects the $f$ electrons to form a half-filled antiferromagnet.

The simple physical picture leading to the effective Hamiltonian
$H_{\rm eff}$ is that the coupling between the $f$ and $a$ electrons
and the antiferromagnetic ordering of the former produces an effective
staggered magnetic field that couples to the spin of the $a$
electrons.  The staggered magnetic field ${\bf B}$ is determined by
the interband coupling (see below).  This effect is independent of the
filling in the $a$ band, because the antiferromagnetism comes from the
$f$ band, which is half-filled.  This makes the situation simpler than
in the non-half-filled repulsive Hubbard model, where commensurate
order cannot be assumed.

In the remainder of this section, we derive $H_{\rm eff}$ from $H$ by
an argument patterned after Anderson's derivation of the Heisenberg
antiferromagnet from the half-filled Hubbard model.

We do a second-order perturbation expansion in $\lambda $ to show that
the coupling between the $a$ and the $f$ electrons produces the
staggered field term of the effective Hamiltonian.  For $\lambda =0$,
the ground state wave function factors into the product of that of the
ground state for the $a$ and the $f$.  Since the $f$ system is
half-filled because $U$ is large, the ground state wave function for
the $f$ electrons is
\eq
\Neel = \prod\limits_{{\bf x}} f^+_{\sigma({\bf x}) }({\bf x})
|0\rangle\, ,\quad \sigma({\bf x}) = e^{{i} {\bf Q} \cdot {\bf x} }
\eqend
where $|0\rangle$ denotes the empty state (and $\sigma({\bf x})$ means
$\uparrow $ if $+1$ and $\downarrow $ otherwise). The Hilbert space of
the total system is the sum of the two orthogonal subspaces ${\cal
H}_a$ and ${\cal H}_f$. Correspondingly, the Hamiltonian has the
matrix structure ($V_{fa}=\lambda V$ and $H_a=H_{\rm hop}+H_{\rm int}$)
\eq
H = \left( \matrix{ H_a & V_{fa}\cr
                    V_{fa}& H_f \cr}\right)
= H_0 + \left(\matrix{0&\lambda V\cr \lambda V&0\cr}\right) \, .
\eqend
The potential term appears off-diagonal because it couples the two
subspaces. The resolvent is also a matrix, and the resolvent equation
\neqa
\f{1}{E-H} &=& \f{1}{E-H_0} +
\f{1}{E-H_0} \ V_{fa} \ \f{1}{E-H}
\\
&=& \f{1}{E-H_0}+
\f{1}{E-H_0} \ V_{fa} \ \f{1}{E-H_0}
\\
&+& \f{1}{E-H_0} \ V_{fa} \ \f{1}{E-H_0} \ V_{fa}\ \f{1}{E-H}
\neqaend
reads, denoting $R_a=(E-H_a)^{-1}$ and $R_f=(E-H_f)^{-1}$,
\eqa
\f{1}{E-H} &=& \left( \matrix{R_a & 0 \cr 0 & R_f\cr}\right) +
\lambda \left(\matrix{0&R_a V R_f \cr R_f V R_a &0\cr}\right) \nonu\nonu
&+& \lambda^2
\left(\matrix{R_a V R_f V & 0\cr 0 & R_f V R_a V\cr}\right)\ \f{1}{E-H}
\, .
\eqaend
We want to calculate the effective propagation of the $a$ electrons to
second order in $\lambda $. For this we need only the left upper
element $R_{\rm eff}=\left(\f{1}{E-H}\right)_{aa}$.  It is given by
\eq
R_{\rm eff}=R_a + \lambda^2 R_a V R_f V R_{\rm eff}
\eqend
or in other words
\eq
{R_{\rm eff}}^{-1}= E-H_a - \lambda^2 V R_f V
\eqend
Up to this point, everything was exact. Now we approximate by keeping
only terms up to order $\lambda^2$. Moreover, since we want to see the
influence of the interaction in the ground state only, we may
calculate $V R_f V$ by its expectation with respect to $f$, taken in
the state $\Neel$. This gives a staggered magnetic field term
\eq
\f{1}{E+\epsilon_f+U}
\sum_{{\bf x}} a^+({\bf x})\sigma_3 a({\bf x}) \, e^{{i} {\bf Q} \cdot {\bf x} }
\eqend
Finally, we want to define the effective Hamiltonian $H_{\rm eff}$ by
${R_{\rm eff}}^{-1}=E-H_{\rm eff}$. To do so, we note that the energy
$E$ is in the $a$ band, so $|E|\le 2t << \epsilon_f+U$, and thus we
may replace $\f{1}{E+\epsilon_f+U}$ by $\f{1}{\epsilon_f+U}$. With
this approximation, ${R_{\rm eff}}^{-1}$ depends only linearly on $E$,
and, up to an uninteresting shift in the energy,
\eq
H_{\rm eff} = H_a + \f{\lambda^2}{\epsilon_f+U}
\sum_{{\bf x}}a^+({\bf x})\sigma_3  a({\bf x}) \, e^{{i} {\bf Q} \cdot {\bf x} }
\eqend
is the effective Hamiltonian.  Since $H_{\rm int}$ couples only the
$a$ electrons, it does not play a role in this argument. The coupling
also produces a feedback on the $f$ electrons, but this influences
the $a$ electrons only in higher order in $\lambda$. This shows that a
model of the type we study is induced, but the ${\bf B}$ field obtained by
this argument is small.

\section{Formalism}\label{formalism}

\subsection{Functional integrals and mean field approximation}

To study our effective model, we use the path integral formalism
\cite{NO}. It is equivalent to the Hamiltonian framework and more
convenient for calculating the grand canonical partition function
$Z={\rm Tr}\,e^{-\beta(H-\mu N)}$, which is thus expressed as a
functional integral $Z=\int{\cal D}\bar\psi {\cal D}\psi \, e^{-{\cal
A}}$ over Grassmann variables $\psi_{\sigma}(x)$ and
${\bar\psi}_{\sigma}(x)$ carrying a spin index $\sigma = \{ \uparrow ,
\downarrow \}$.  Configuration space is labeled by $x=(\tau_x,{\bf
x})$ where $0 < \tau_x < \beta$ is the usual imaginary time and ${\bf
x}$ are vectors in the $d$-dimensional cubic lattice $\Z^d$ (i.e.\ we
set the lattice constant equal to 1).  Most of our calculations are
for arbitrary $d$, but we are mainly interested in $d=2$ or $3$.  In
intermediate steps of our derivation we will implicitly assume that
space is a finite square (cube) with a finite number $L^d$ of points,
but we will eventually take the thermodynamic limit $L \to\infty$.  We
group the spacetime argument $x$ and the spin index $\sigma$ to a
single coordinate $X=(x,\sigma)$, and set $\psi(X)=\psi_\sigma (x)$
etc.  We also use convenient short hand notations
$\sum_x\equiv\int_0^{\beta}d\tau_x \sum_{{\bf x}\in\Lambda}$ and
$\sum_X = \sum_{x,\sigma}$ etc.

The action of the model is ${\cal A} = \sum_{X} \bar\psi(X)
(\partial_\tau-\mu) \psi(X) + \int_0^\beta H(\bar\psi,\psi)d\tau
={\cal A}_2 + {\cal A}_4$, with a quadratic part defining the free
Green function, ${\cal A}_2 = -\sum_{X,Y}\bar\psi(X) {G_0}^{-1}(X,Y)
\psi(Y)$, and a quartic interaction term.

To derive mean field equations for possible superconducting states we
now introduce auxiliary bilocal pair fields and thus decouple the
fermion interaction (see e.g.\ \cite{Kleinert}),
\eqa
\label{FHF}
e^{-{\cal A}_4} = \int {\cal D}\Delta^* {\cal D}\Delta \,
\exp \sum_{X,Y}\Bigl(\f{|\Delta(X,Y)|^2}{2 V(X,Y)} \nonu
-
\half\overline{\Delta(X,Y)}\psi(Y)\psi(X) -
\half\bar\psi(X)\bar\psi(Y)\Delta(X,Y)
\Bigr)
\eqaend
where
\eq
{\cal D}\Delta^*{\cal D}\Delta = \prod\limits_{\{X,Y\}}
\frac{d{\rm
Re}\Delta(X,Y) d{\rm Im}\Delta(X,Y)}{2 \pi |V(X,Y)|}
\eqend
is the measure for a complex boson path integral. The product runs
only over those unordered pairs $\{ X,Y\}$ for which $V(X,Y) \ne 0$.
We call the $\Delta(X,Y)$ Hubbard-Stratonovich (HS) fields.  Note
that the integral over $\Delta(X,Y)$ is convergent because (and only
if) the pairing potential is purely attractive, i.e.\ $V(X,Y)< 0$
whenever it is nonzero.  Note also that due to the anticommutativity
of the Grassmann variables only antisymmetric configurations,
\eq
\label{pauli}
\Delta(X,Y) = -\Delta(Y,X)\, ,
\eqend
contribute to the path integral. This corresponds to the Pauli principle.

Integrating out the fermions now leads to
\eq
\label{HS}
Z = \int {\cal D}\Delta^* {\cal D}\Delta \, e^{ -{\cal F}(\Delta) }
\eqend
with the effective HS action
\eq
\label{F}
{\cal F}(\Delta) =
-\sum_{X,Y}\f{|\Delta(X,Y)|^2}{2 V(X,Y)} -\half {\rm Tr}\log({\cal G}^{-1})
\eqend
where the second term come from the fermion path integral (logarithm
of Pfaffian of ${\cal G}^{-1}$ = half of logarithm of determinant of
${\cal G}^{-1}$), and we have introduced
\eq
\label{Dyson}
{\cal G}^{-1} = {\cal G}_0^{-1} -\Sigma
\eqend
with
\eq
\label{nG}
{\cal G}_0=\left(\matrix{G_0&0\cr 0& \tilde{G_0}}\right), \quad
\tilde{G_0}(X,Y)= -G_0(Y,X)
\eqend
the free Nambu Green function and
\eq
\label{Sigma}
\Sigma = \left(\matrix{ 0 &\Delta \cr \Delta^* & 0}\right),
\quad
\Delta^*(X,Y)= \overline{\Delta(Y,X)} \: .
\eqend

In the following we write
\eq
\label{Nambu}
{\cal G} =
\left(\matrix{G &F\cr F^* &\tilde G}\right)
\eqend
which is equal to the Nambu Green function for non-interacting
fermions in an external field $\Delta$.

HF theory amounts to evaluating the HS path integral in
Eq.\ (\ref{HS}) using the saddle point method.  We obtain $Z \approx
\exp[-{\cal F}(\Delta_{HF}) ]$ where $\Delta_{HF}$ is the solution of
the saddle point equation $\delta{\cal F}(\Delta)/\delta\Delta(X,Y)=0$
which corresponds to the {\em minimum} of ${\cal F}$.  More explicitly
the (complex conjugate of the) latter equation is
\eq
\label{MF}
\Delta(X,Y) = V(X,Y) F(X,Y)
\eqend
which together with Eq.\ (\ref{Dyson}) forms a self consistent system
of equations and provides the BCS description for our model.  Note
that in the saddle--point approximation, $-\langle \Psi(X)
\Psi(Y)^\dagger\rangle = {\cal G}$, so Eq.\ (\ref{Dyson}) can now be
interpreted as the Dyson equation for our electron Green function
${\cal G}$ where mean field theory gives the approximation (\ref{MF})
for the electron self energy $\Sigma$.

The general analysis of the mean field equations is still too
difficult since for general $X,Y$ dependent HS field configurations
$\Delta(X,Y)$, the evaluation of the Nambu Green function
(\ref{Nambu}) is impossible in general.  To proceed we therefore make
the usual assumption and only consider HS configurations that have at
least a remnant of translation invariance.  Since for our model the
free Green function contains a staggered contribution i.e.\ is not
invariant under translations by {\em one} but only {\em two} sites,
the simplest consistent ansatz for our HF equations are HS
configurations $\Delta$ invariant by translations by two sites.

\section{Staggered superconductivity}\label{MFeqs}

\subsection{Staggered states}

We now specialize to time independent states where the superconducting
order parameter has a uniform part and a staggered part,
i.e. \cite{note}
\eq
\label{staggered}
\Delta({\bf x},{\bf y}) = \Delta_N({\bf x}-{\bf y}) +
\Delta_A({\bf x}-{\bf y}) e^{i{\bf Q \cdot y}},
\eqend
with ${\bf Q} = (\pi,\ldots,\pi)$ the antiferromagnetic vector (here
and below we use spin matrix notation whenever possible).  This
obviously implies that ${\cal G}$ is also a sum of a uniform and
staggered contribution, and we can solve the Dyson equation
(\ref{Dyson}) by Fourier transformation. We use the convention
\[ f(x) = \sum_k e^{-{i} kx} f(k)\, ,
\quad kx=\omega_n\tau_x - {\bf x}\cdot{\bf k} \]
where $k=(\omega_n,{\bf k})$ with $\omega_n=\frac{(2n+1)\pi}{\beta}$
($n\in\Z$) the Matsubara frequencies and ${\bf k}=(k_1,\ldots, k_d)$ a
momentum vector in the first Brillouin zone $[-\pi,\pi]^d$ of the
lattice, i.e.\
\[
\sum_k \equiv \frac{1}{\beta}\sum_{\omega_n}\intk\, ,\quad
\intk\equiv \int\limits_{[-\pi,\pi]^d}
\frac{d^d {\bf k}}{(2\pi)^d}\, .
\]
Then $f(k) = \sum_x e^{{i} kx} f(x)$.

With that the self-energy equation (\ref{MF}) becomes
\eq \Delta_{N,A}(k) = \sum_q V(k-q)F_{N,A}(q) \, .
\label{MF2}
\eqend

\subsection{$T_c$-equations}
\label{Tceqs}

We first consider the region close to the critical temperature $T_c$
where the Dyson equation (\ref{Dyson}) can be linearized in the
non-gauge invariant Green functions $F$ and self energy $\Delta$.
Writing (\ref{Dyson}) as ${\cal G} = {\cal G}_0 + {\cal
G}_0*\Sigma*{\cal G}_0 +\cdots$ we obtain the linearized equation $F=
G_0*\Delta*\tilde G_0$ i.e.\
\eq
F(x,y) = \sum_{z_1,z_2}G_0(x,z_1)\Delta(z_1,z_2)\tilde{G}_0(z_2,y)
\eqend
where $\tilde G_0(x,y) \equiv -G_0(y,x)^T$ and $^T$ means matrix
transposition.  With the ansatz (\ref{staggered}) and Fourier
transform this becomes (the $k$ dependence is suppressed, $F_N\equiv
F_N(k)$ etc., in the following),
\eqa\label{FNAdef}
F_{N,A} = G_{0N} \Delta_N \tilde{G}_{0N,A} +
 G_{0A} \Delta_N^Q\tilde{G}_{0A,N}^Q \nonu
 +G_{0A} \Delta_A^Q\tilde{G}_{0N,A} + G_{0N} \Delta_A\tilde{G}_{0A,N}^Q
\eqaend
where $\Delta_N^Q(k)= \Delta_N(k-Q)$ etc, and spin matrix
multiplication is understood.  Combined with Eq.~(\ref{MF2}), this
gives the $T_c$ equations for our model.

The translation-invariant and staggered parts of the free Green
function, $G_{0N,A}$ are
\begin{mathletters}
\label{Gn}
\eq
G_{0N}(k) = g(k)\sigma_0,\quad G_{0A}(k) = a(k) \sigma_3
\eqend
where
\eqa
g(k) = \f{{i}\omega_n +\mu + \epsilon({\bf k})}{({i}\omega_n - E_+
)({i}\omega_n -E_- )}\nonu
a(k) =  \f{{s}}{({i}\omega_n - E_+ ) ({i}\omega_n - E_-)}
\eqaend
\end{mathletters}
[we used $\epsilon({\bf k} + {\bf Q}) = -\epsilon({\bf k})$] and
\eq
\label{Epm}
E_\pm({\bf k}) = -\mu \pm \sqrt{\epsilon({\bf k})^2 + s^2}
\eqend
are the antiferromagnetic bands separated by a gap $=2s$. We also
set ${\bf B \cdot \mbox{\boldmath$\sigma$}} = s\sigma_3$ without loss
of generality.

\begin{table}
\caption{Symmetries of the order parameter.}
\label{etaJ}
\begin{tabular}{c|rrr}
Symmetry & On-site (0) & Nearest neighbor (nn) & Next nearest
neighbor (nnn) \\ \hline
$s$-wave &
$\eta_0({\bf k})  = 1$ &
$\eta_{1}({\bf k}) = \cos(k_1)+ \cos(k_2)$ &
$\eta_{5}({\bf k})  = \cos(k_1+k_2) + \cos(k_1-k_2)$ \\
$p$-wave &
&
$\eta_{2,3}({\bf k})  = \sin(k_1)\pm \sin(k_2)$ &
$\eta_{6,7}({\bf k}) = \sin(k_1+k_2)\pm \sin(k_1-k_2)$ \\
$d$-wave &
&
$\eta_{4}({\bf k}) = \cos(k_1) -  \cos(k_2)$ &
$\eta_{8}({\bf k})  = \cos(k_1+k_2) - \cos(k_1-k_2)$
\end{tabular}
\end{table}

To deal with the remaining dependence on the relative coordinate,
${\bf x - y}$, we decompose the order parameter $\Delta_{N,A}$ in
components that transform according to the irreducible representations
of the lattice point group:
\eq
\label{decomp}
\Delta({\bf x}-{\bf y}) = \sum_J \Delta_J \eta_J({\bf x}-{\bf y}).
\eqend
For a two dimensional square lattice the functions $\eta_J$ we need
are listed in Table~\ref{etaJ}.  They are orthonormal, $\intk
\eta_J({\bf k})\eta_{J'}({\bf k})=\delta_{JJ'}$.  These functions
determine the shape of the gap, i.e., if it is $s$-wave or $d$-wave
etc.  The potential (\ref{pot}) has the expansion
\eq
\label{effpot}
V({\bf k}-{\bf q}) = -\sum_{J} g_J \eta_J({\bf k})\eta_J({\bf q}).
\eqend
where
\eqa
g_{1} &=& g_{2} = g_{3} = g_{4} = g_{\rm nn} \nonu
g_{5} &=& g_{6} = g_{7} = g_{8} = g_{\rm nnn}.
\eqaend
Note that the interaction $V$, which is a diagonal matrix in the
$x$-basis (i.e., it acts by ordinary multiplication) in
Eq.~(\ref{MF}), is still diagonal in the new basis.  This is because
the coupling constants $g_0$, $g_{\rm nn}$ and $g_{\rm nnn}$ are the same for
all on-site, for all nearest neighbor and for all next nearest
neighbor interactions, respectively, i.e., the interaction is
proportional to identity within each of these subspaces, and because
the unitary transformation implied by Eq.~(\ref{decomp}) does not mix
these subspaces.
Then we get coupled (linear algebraic) equations for the $2\times 2$
matrices $(\Delta_{N,A})_J$.  The different channels $J$ do not mix in
the $T_c$-equations:
taking for instance the nn interaction, we have $V({\bf
k}-{\bf q}) = -g \sum\limits_{J=1}^4 \eta_J({\bf k}) \eta_J({\bf q})$.
Inserting Eq.~(\ref{effpot}), Eq.~(\ref{FNAdef}), and Eq.~(\ref{decomp})
into Eq.~(\ref{MF2}), we get a system of equations whose coefficients
contain integrals of the form $\intk \eta_J({\bf k}) \eta_{J'} ({\bf k})
\Phi ({\bf k})$, where $\Phi$ is a function that has the symmetries $\Phi
(-{\bf k}) =
\Phi({\bf k})$ and $\Phi(k_2,k_1)=\Phi(k_1,k_2)$ because it depends on ${\bf k}$
only through $\epsilon({\bf k})$.  The first symmetry implies that $p$-wave
($J=2,3$) does not couple to $s$ and $d$--wave ($J=1,4$), and the second
symmetry implies that $s$ and $d$ are uncoupled.

We write
\eqa
\label{approx}
\Delta_{N}(k) =\left(\matrix{\iDelta_{1,J} & \iDelta_{2,J}\cr \iDelta_{3,J}
&\iDelta_{4,J}
}\right)\eta_J({\bf k})
\nonu
\Delta_{A}(k) =\left(\matrix{\iTheta_{1,J} &
\iTheta_{2,J}\cr \iTheta_{3,J} &\iTheta_{4,J} }\right)\eta_J({\bf k})
\eqaend
so that the $\iDelta_{i,J}$ and $\iTheta_{i,J}$ are the c-number
order-parameters for the superconducting state.  The final equations
will give a critical temperature $T_c$ for each of the possible
channels separately, and the biggest of those determines the
dominating channel and the $T_c$ of the system.

{}From now on, we suppress the indices $(J,i)$ and write \[\eta
\equiv\eta_J,\quad g_J\equiv \gg\] etc.; it is clear from the above
that one has to calculate the critical temperatures for all
interesting channels and take the maximum one.  This determines the
dominant SC channel: the `shape' $\eta({\bf k})$, the spin structure,
and the ratio of translation invariant and staggered parts $\iDelta$
and $\iTheta$ of the gap.

Every shape function $\eta$ can be characterized by two sign factors
$p=+,-$ and $\xi=+,-$ defined by
\eq
\eta({\bf k}-{\bf Q}) = p\, \eta({\bf k})\, , \quad \eta(-{\bf k}) =
\xi\eta({\bf k})\, ;
\eqend
these determine which spin structures of the gap are compatible with
the Pauli principle Eq.\ (\ref{pauli}). Since in position space, the
spatial dependence of $\Delta_{N,A}({\bf x}-{\bf y})$ is given by the
Fourier transform $\eta({\bf x}-{\bf y}) = p\, \eta({\bf x}-{\bf y})
e^{{i}{\bf Q}\cdot ({\bf x}-{\bf y})} =\xi \eta({\bf y}-{\bf x}) $ of
$\eta({\bf k})$, Eq.\ (\ref{pauli}) implies the following conditions
\begin{mathletters}
\label{pauli1}
\eq
\iDelta_{1,4} = -\xi\iDelta_{1,4} \, ,\quad \iTheta_{1,4} = -\xi p\,
\iTheta_{1,4}
\eqend
and
\eq
\iDelta_2 = -\xi\iDelta_3\, ,\quad \iTheta_2 = -\xi p\, \iTheta_3 \: .
\eqend
\end{mathletters}
These conditions exclude many of the possible solutions of the mean
field equations, e.g.\ all those where $\iDelta_{1}\neq 0$ with
$\xi\neq -1$ and $\iTheta_{1}\neq 0$ with $\xi p\neq -1$ etc.

We now derive the $T_c$-equations for our model.  As the potential is
frequency independent we can do the Matsubara sums analytically.
After a straightforward calculation (see Appendix~\ref{AppB}) we
obtain
\begin{mathletters}
\label{Tc}
\eqa
\iDelta_{1,4} = \gg K_p(\beta)\iDelta_{1,4}\pm\gg
L_p(\beta)\iTheta_{1,4}\nonu
\iTheta_{1,4} = \gg \tilde K_p(\beta)\iTheta_{1,4} \pm
\gg L_p(\beta)\iDelta_{1,4}
\eqaend
and
\eqa
\iDelta_{2,3} = \gg K_{-p}(\beta)\iDelta_{2,3} \mp
  \gg L_{-p}(\beta)\iTheta_{2,3} \nonu
\iTheta_{2,3} = \gg\tilde K_{-p}(\beta)\iTheta_{2,3} \mp
  \gg L_{-p}(\beta)\iDelta_{2,3}
\eqaend
\end{mathletters}
where $p=\pm$ is determined by $\eta({\bf k})$ as above, and the
functions $K_+(\beta)$ etc.\ are given in Appendix~\ref{AppB}.  In
fact $L_-=0$, therefore half of the combinations will lead to
uncoupled equations for the staggered and translation invariant gaps.

In Table~\ref{channels} we list the possible channels
$\iDelta({\bf x},{\bf y})$.  We use the following notation for the
spin matrix structure
\eq
|\!\uparrow\uparrow\rangle = \left(\matrix{ 1&0\cr 0&0 }\right)  \, ,\quad
|\!\uparrow\downarrow\rangle= \left(\matrix{ 0&1\cr 0&0 }\right) \, ,
\quad \mbox{
etc.}
\eqend
and $\eta_J$ is short for $\eta_J({\bf x}-{\bf y})\equiv \intk\,
e^{{i}{\bf k}\cdot({\bf x-y}) }\eta({\bf k})$ i.e.\ the Fourier
transformed functions defined in Table~\ref{etaJ}.  We use obvious
symbols $\iDelta_{s}^0$ etc.\ as abbreviations for the channels.

\begin{table}
\caption{Possible channels $\iDelta({\bf x},{\bf y})$.}
\label{channels}
\begin{tabular}{cc}
On-site ( $p=+1$ )\\ \hline
$\iDelta_{s}^0$ & $ \iDelta\, \bigl(
|\!\uparrow\downarrow\rangle -  |\!\downarrow\uparrow\rangle \bigr)
\, \eta_0$ \\

$\iTheta_{s}^0$ & $\iTheta\,
e^{{i}{\bf Q}\cdot {\bf y} }  \bigl(
|\!\uparrow\downarrow\rangle -  |\!\downarrow\uparrow\rangle \bigr)
\, \eta_0$
\\ \hline
Nearest neighbor ( $p=-1$ )\\ \hline

$\iTheta^{\rm nn}_{s}$ & $ \iTheta\,
e^{{i}{\bf Q}\cdot {\bf y}} \bigl\{ |\!\uparrow\uparrow\rangle
\mbox{ or } |\!\downarrow\downarrow\rangle  \bigr\} \, \eta_1$ \\

$\iDelta^{\rm nn}_{p }$&$\:\: \iDelta\, \bigl\{ |\!\uparrow\uparrow\rangle
\mbox{ or } |\!\downarrow\downarrow\rangle  \bigr\} \, \eta_{2,3}$\\

$\iTheta^{\rm nn}_{d}$&$\:\: \iTheta\,
e^{{i}{\bf Q}\cdot {\bf y}} \bigl\{ |\!\uparrow\uparrow\rangle
\mbox{ or } |\!\downarrow\downarrow\rangle  \bigr\} \, \eta_4$\\

$(\iDelta\&\iTheta)^{\rm nn}_{s}$&
$\Bigl( \iDelta \bigl(|\!\uparrow\downarrow\rangle -
|\!\downarrow\uparrow\rangle \bigr) +
\iTheta\, e^{{i}{\bf Q}\cdot{\bf y}} \bigl(|\!\uparrow\downarrow\rangle +
|\!\downarrow\uparrow\rangle \bigr) \Bigr) \, \eta_1$\\

$(\iDelta\&\iTheta)^{\rm nn}_{p}$&
$\Bigl( \iDelta \bigl(|\!\uparrow\downarrow\rangle +
|\!\downarrow\uparrow\rangle \bigr) +
\iTheta\, e^{{i}{\bf Q}\cdot{\bf y}} \bigl(|\!\uparrow\downarrow\rangle -
|\!\downarrow\uparrow\rangle \bigr) \Bigr) \, \eta_{2,3}$\\
$(\iDelta\&\iTheta)^{\rm nn}_{d}$&
$\Bigl( \iDelta \bigl(|\!\uparrow\downarrow\rangle -
|\!\downarrow\uparrow\rangle \bigr) +
\iTheta\, e^{{i}{\bf Q}\cdot{\bf y}} \bigl(|\!\uparrow\downarrow\rangle +
|\!\downarrow\uparrow\rangle \bigr) \Bigr) \, \eta_4
$\\ \hline
Next nearest neighbors ( $p=+1$ ) \\ \hline
$\iDelta_{s}^{\rm nnn}$ & $\iDelta\,  \bigl(
|\!\uparrow\downarrow\rangle -  |\!\downarrow\uparrow\rangle \bigr)
\, \eta_5$\\

$\iDelta_{p}^{\rm nnn}$ & $ \iDelta\,  \bigl(
|\!\uparrow\downarrow\rangle +  |\!\downarrow\uparrow\rangle \bigr)
\, \eta_{6,7}$\\
$\iDelta_{d}^{\rm nnn}$ &$ \iDelta\,  \bigl(
|\!\uparrow\downarrow\rangle -  |\!\downarrow\uparrow\rangle \bigr)
\, \eta_8$\\
$\iTheta_{s}^{\rm nnn}$ & $ \iTheta\,
e^{{i}{\bf Q}\cdot {\bf y} }  \bigl(
|\!\uparrow\downarrow\rangle -  |\!\downarrow\uparrow\rangle \bigr)
\, \eta_5$\\

$\iTheta_{p}^{\rm nnn}$ & $ \iTheta\,
e^{{i}{\bf Q}\cdot {\bf y} }  \bigl(
|\!\uparrow\downarrow\rangle +  |\!\downarrow\uparrow\rangle \bigr)
\, \eta_{6,7}$\\
$\iTheta_{d}^{\rm nnn}$ & $ \iTheta\,
e^{{i}{\bf Q}\cdot {\bf y} }  \bigl(
|\!\uparrow\downarrow\rangle -  |\!\downarrow\uparrow\rangle \bigr)
\, \eta_8$\\
$(\iDelta\&\iTheta)^{\rm nnn}_{p}$ & $ \left\{  \matrix{
\bigl( \iDelta + \iTheta\, e^{{i}{\bf Q}\cdot{\bf y}}  \bigr)\,
|\!\uparrow\uparrow\rangle\, \eta_6
  \cr
\bigl( \iDelta - \iTheta\, e^{{i}{\bf Q}\cdot{\bf y}}  \bigr)\,
|\!\downarrow\downarrow\rangle\, \eta_7} \right.
$\\
\end{tabular}
\end{table}

Note that some of the gaps are a mixture of spin singlet and
triplet. This comes as no surprise since spin rotation symmetry is
broken in our model.

We will analyze these $T_c$-equations numerically in the next section.
First, however, we discuss a simplified analysis which applies in the
limit where the pairing across the AF gap can be neglected:
For large $s$ and large $\beta$ (small temperatures), a nontrivial
filling $\rho\neq 1$ is only possible for $s\approx w\approx |\mu|$,
thus \eq |K_+| \approx |\tilde K_+| \approx |L_+| \approx M
\log(\Omega\beta)\, , \eqend and $K_-,\tilde K_-$ are much smaller in
comparison (explicit formulas for these constants are derived in
Appendix~\ref{AppB}).  The constant 
\eq
\label{M}
M  = \intk \f{1}{2}\bigl(\delta(E_+) +\delta(E_-)\bigr)\eta({\bf k})^2
\eqend
has the natural interpretation as DOS $N_s(\mu)$ times the Fermi
surface average of $\eta({\bf k})^2$ (shape of the SC gap-squared),
and $\Omega$ can be regarded as an effective energy cutoff of the
interaction.  Thus we get a BCS like equation
\eq
\label{bcs}
T_c=\Omega\, e^{-1/\gg \Lambda}
\eqend
with $\Lambda=2M$, which can be trusted as long as $T_c$ is small.
This already allows us to make predictions about the most stable
channels for large $s$-values: the channels where $\iDelta$ and
$\iTheta$ are decoupled should have a negligible $T_c$ (due to the
smallness of $K_-$ and $\tilde K_-$) and thus should be irrelevant as
compared to the mixed channels $(\iDelta\&\iTheta)_{\cdots}^{\cdots}$.
Moreover, for the latter we should always find $\iTheta \approx \pm
\iDelta$, at least for sufficiently large $s$. Thus in case of dominating
nnn attraction one should expect the SC channel
$(\iDelta\&\iTheta)^{\rm nnn}_{p}$.  In case of dominating nn
attraction, also the shape of the gap matters, and one has to check
which one leads to the biggest $M$ i.e.\ the largest Fermi surface
average of $\eta({\bf k})^2$.  For larger $s$-values, the Fermi
surface is similar to the one of the half-filled hopping band (since
the Fermi surface is given by $E_\pm = 0$ i.e.\ $\epsilon({\bf k}) =
\pm \sqrt{\mu^2-s^2}$ which is $\approx 0$ for larger $s$-values at
nonzero doping), and $\eta_4$ leads to the biggest $M$ (it is easy to
see why $\eta_1$ cannot lead to a significant $M$: it is proportional
to $\epsilon$ and thus $\approx 0$ at the Fermi surface).  Thus for
dominating nn attraction, one should expect the SC channel
$(\iDelta\&\iTheta)^{\rm nn}_{d}$.  We confirm this result in a
systematic stability analysis of all channels in the next section.

The $\iDelta\&\iTheta$-structure of the dominating SC channel has a
simple physical explanation: The density of electrons should follow
the staggered magnetic field ${\bf B}e^{{i}{\bf Q}\cdot{\bf x}}$:
denoting as $\uparrow$ the spin directions in the direction of ${\bf
B}$ and $\downarrow$ opposite to it, the density difference
$\rho_\uparrow({\bf x})-\rho_\downarrow({\bf x})$ of spin-$\uparrow$
and spin-$\downarrow$ fermions on the site ${\bf x}$ should be
proportional to $|{\bf B}|e^{{i}{\bf Q}\cdot{\bf x}}$.  Thus one
should expect that stable Cooper pairs can only arise from electrons
with their spins in direction of ${\bf B}e^{{i}{\bf Q}\cdot{\bf x}}$
(otherwise the density of the participating electrons is very small).
If both electrons are on the same site (on-site attraction), this is
never possible and no significant SC is expected.  In case the paired
electrons are on adjacent sites (nn attraction), the preferred spin
configuration is expected to be $|\!\uparrow\downarrow\rangle$ or
$|\!\downarrow\uparrow\rangle$ (following the AF background), and there
should be nearly no mixing of these two.  Finally in the nnn case, the
most stable Cooper pairs should be those with
$|\!\uparrow\uparrow\rangle$ or $|\!\downarrow\downarrow\rangle$
(following the AF background).

\subsection{Mean field theory below $T_c$}\label{btc}

The mean field theory below $T_c$ is obtained by minimizing the
action, Eq.~(\ref{F}), for field configurations of the form given in
Eq.~(\ref{staggered}).  Introducing also the transformation
(\ref{decomp}) we can write the action as
\eq
{\cal F}= \beta L^d \sum_{J,\sigma,\sigma'}
\f{|\iDelta_{J\sigma\sigma'}|^2+|\iTheta_{J\sigma\sigma'}|^2}{2g_J} -
\half {\rm Tr} \log \left({\cal G}^{-1}\right) \, .
\eqend

The second term in the action can be expressed as ${\rm Tr} \log
\left({\cal G}^{-1}\right)=\half \beta L^d \sum_k \log\det{\cal
R}(k)$, where ${\cal R}$ is the $8\times 8$-matrix given by
\eq
{\cal R}=\left(\matrix{R & -\Delta \cr -\Delta^* & \tilde R } \right)
\eqend
where $\tilde R(k) = - R^T(-k)$ with
\eq
R = \left( \matrix{R_N & R_A \cr R_A^Q & R_N^Q } \right) \; , \quad
\Delta = \left( \matrix{ \Delta_N   & \Delta_A \cr
                         \Delta_A^Q &  \Delta_N^Q } \right)
\eqend
and
\eq
\label{QNA}
R_N = [ i\omega_n - (\epsilon({\bf k}) - \mu) ] \sigma_0 \; , \quad
R_A = - s \sigma_3 \; .
\eqend
We assume that only one channel will contribute, i.e., we put all
$g_J$ but one equal to zero.  In contrast to the $T_c$-equation, this
absence of channel mixing is not enforced by a symmetry but is an
additional assumption.  Then, with Eqs.~(\ref{approx}), (\ref{pauli1})
and $\epsilon({\bf k}) = \epsilon(-{\bf k}) = -\epsilon({\bf k-Q})$,
the determinant can be evaluated by finding the eigenvalues $E_c$ of
${\cal R}$:
\eq
\half \log \det{\cal R}(k) = \sum_{c=1}^4 \log\bigl(i\omega -
E_c(k)\bigr)
\eqend
(since every eigenvalue is 2--fold degenerate).  These eigenvalues
describe the electron bands.  For the dominant channels
$\iDelta\&\iTheta$ in Table \ref{channels}, these electron bands
$E_c$ are give by
\begin{mathletters}
\label{bands}
\eq
E_{1,2}=\pm E_+\, \quad E_{3,4}=\pm E_-
\eqend
with
\eqa
E_\pm = \sqrt{\epsilon^2+\mu^2+s^2+|\Delta_N|^2+|\Delta_A|^2\pm 2W  } \nonu
W=\sqrt{\epsilon^2|\Delta_A|^2 + \epsilon^2\mu^2 +[\mu s +
|\Delta_N||\Delta_A|\cos(\phi)]^2 }\nonu
\epsilon=\epsilon({\bf k})\, ,\quad
\Delta_N=\iDelta\eta({\bf k})\, , \quad \Delta_A=\iTheta\eta({\bf k})\,
\eqaend
where $\phi$ is the relative phase of $\iDelta$ and $\iTheta$,
\eq
\label{LRO}
\iDelta = |\iDelta |\, e^{{i} (\alpha + \phi) } \, ,
\quad \iTheta = | \iTheta |\, e^{{i} \alpha}
\eqend
\end{mathletters}
and the bands are independent of $\alpha$.  [We do not write down the
formulas for the other channels since they are
irrelevant, as discussed.]

Performing the Matsubara sum we obtain
\eq
\label{res}
{\cal F} = \beta L^d f_0( \iDelta , \iTheta )
\eqend
with
\eq
\label{f0}
f_0 = \sum_{\sigma,\sigma'}
\f{|\iDelta_{\sigma\sigma'}|^2+|\iTheta_{\sigma\sigma'}|^2}{2g_J} -
\half \sum_{c=1}^4 \intk \f{1}{\beta}\log\bigl(1+e^{-\beta E_c}\bigr) \: .
\eqend
(We note that the Matsubara sum used here was regularized: it is
obtained by integrating
$
\frac{1}{\beta}\sum_{\omega_n} \f{e^{i\omega_n 0^+} }{{i}\omega_n -E} =
\frac{1}{1+e^{\beta E}}
$
and dropping an irrelevant infinite constant.)

The mean field solution is now obtained by minimizing the action,
\begin{mathletters}
\label{MFsol}
\eq
\Omega(\mu)=\min_{\iDelta,\iTheta} f_0 (\iDelta,\iTheta)
\eqend
where we have to take the {\em absolute} minimum.  Moreover, the
chemical potential is fixed by filling,
\eq
\rho=-\f{\partial\Omega}{\partial \mu} \, .
\eqend
\end{mathletters}
The standard mean field equations are $\partial{\cal F}/
\partial\iDelta = \partial{\cal F}/\partial\iTheta = 0$ and have, in
general, several solutions which are not the absolute minimum and
therefore have to be discarded.  In the formulation Eq.~(\ref{MFsol}),
the correct solution is selected automatically.

Clearly the minima will be invariant under global shifts of the phase
$\alpha$.  Thus we end up with an simple minimization problem in the
three parameters $|\iDelta|$,$|\iTheta|$ and $\phi$.  We find that the
relative phase $\phi$ minimizes the action when $W$ is minimized,
i.e.\ $\cos\phi=-{\rm sign}(\mu)$.

With the mean field solution at hand we can now proceed to calculate
physical quantities.  The Green function is given by the matrix
inverse of ${\cal R}(k)$.  As an important example, we recall that the
superconducting one--particle density of states is obtained as
\eq
\label{NE}
N(\omega)=-\f{1}{\pi}{\rm Im}\sum_\sigma \intk
(G_N)_{\sigma\sigma}({\bf k},i\omega_n \rightarrow
\omega + i0^+) \;.
\eqend
Due to the complexity of the bands, (\ref{bands}), this expression is
much more complicated than the standard BCS result.  The differential
current-voltage dependence of the tunneling current between a
superconductor and a normal metal is related to this by
\eq
\label{dIdV}
\f{dI}{dV} \propto \int N(\omega)
\left(-\f{\partial f_\beta(\omega-eV)}{\partial\omega}\right) d\omega \;,
\eqend
where $f_\beta(E)=\f{1}{e^{\beta E}+1}$ is the usual Fermi-Dirac
distribution, and $e$ is the electron charge.


\section{Numerical results} \label{numres}

In this section we describe the results of a numerical evaluation of
the mean field equations in the previous section. The numerical
evaluation was done by performing the Brillouin zone integrals using a
standard integration routine. Since the integrands are typically
strongly peaked, care was taken to ensure sufficiently small
discretization errors by using an adaptive step size method.  Moreover,
we solved mean field equations by minimizing the function $f_0$ Eq.\
(\ref{f0}) using standard numerical routines.

For simplicity we neglect possible mixing of gaps assuming that the
dominating gap transforms under an irreducible representation of the
symmetry group of the lattice (note that below $T_c$, this is an
approximation since in general the energy dependence of the gap cannot
be neglected as in ordinary BCS theory).
We use the following parameters
$
t=0.5\, \mbox{eV}\, , \quad g_{\rm BCS}=0.05\, \mbox{eV}\, ,
\quad 0\leq s\leq 4\, \mbox{eV}
$
motivated by high temperature superconductors \cite{Dagotto} (note
that for the half filled Hubbard model, mean field theory predicts
antiferromagnetic order with $s\approx U/2$).

\begin{figure*}
\bigskip
\centerline{\epsfxsize=9truecm\epsffile{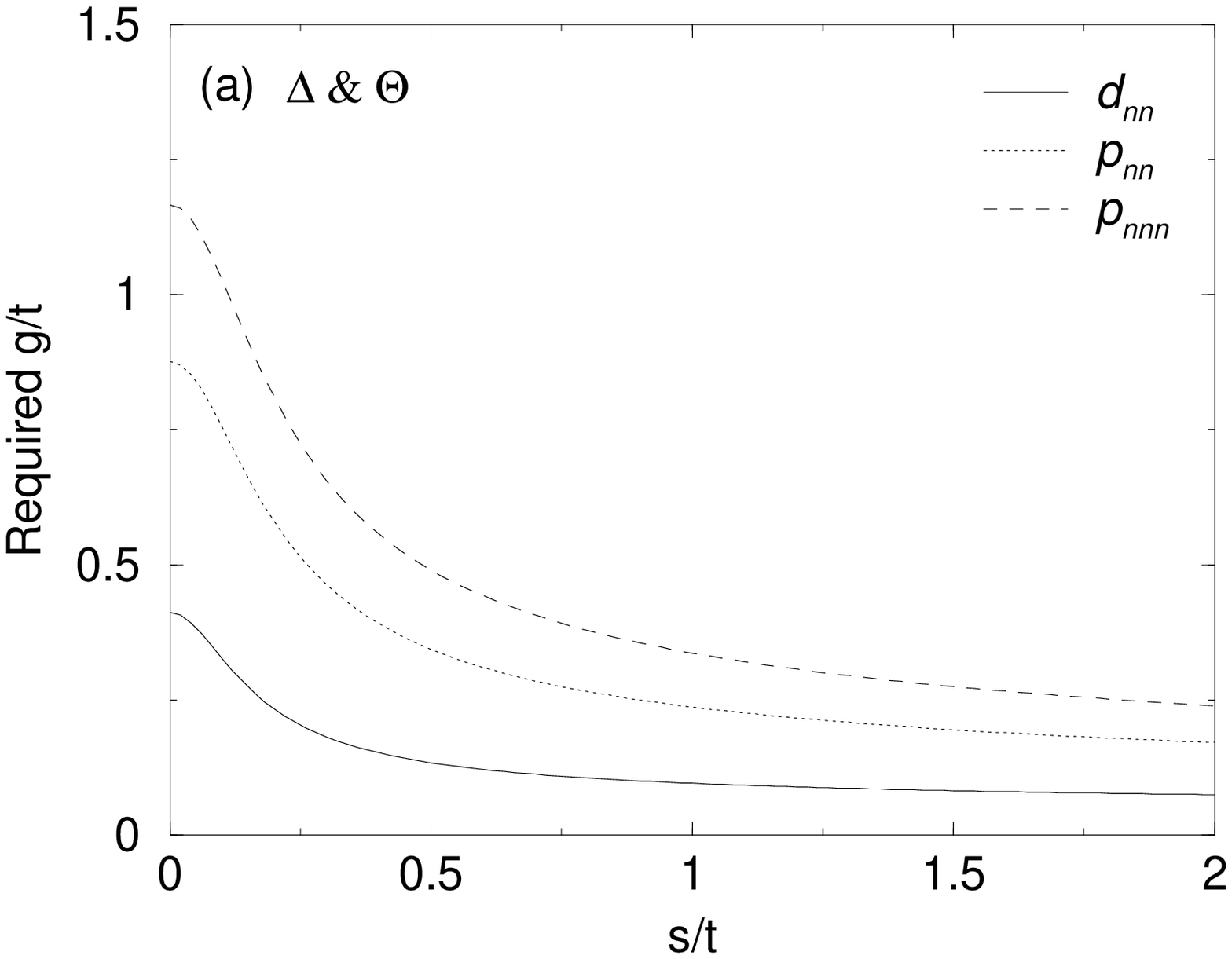}
\epsfxsize=9truecm\epsffile{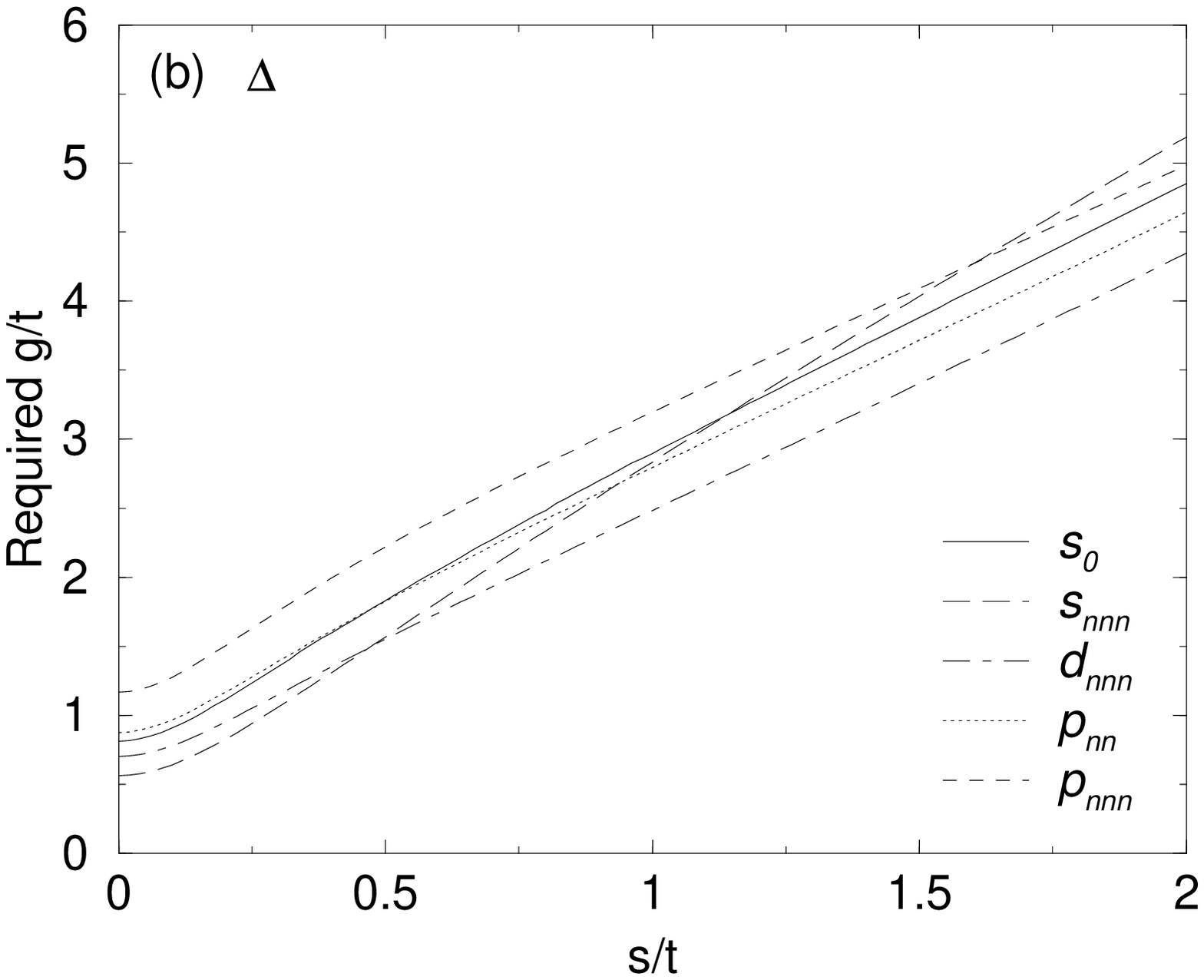}}
\centerline{\epsfxsize=9truecm\epsffile{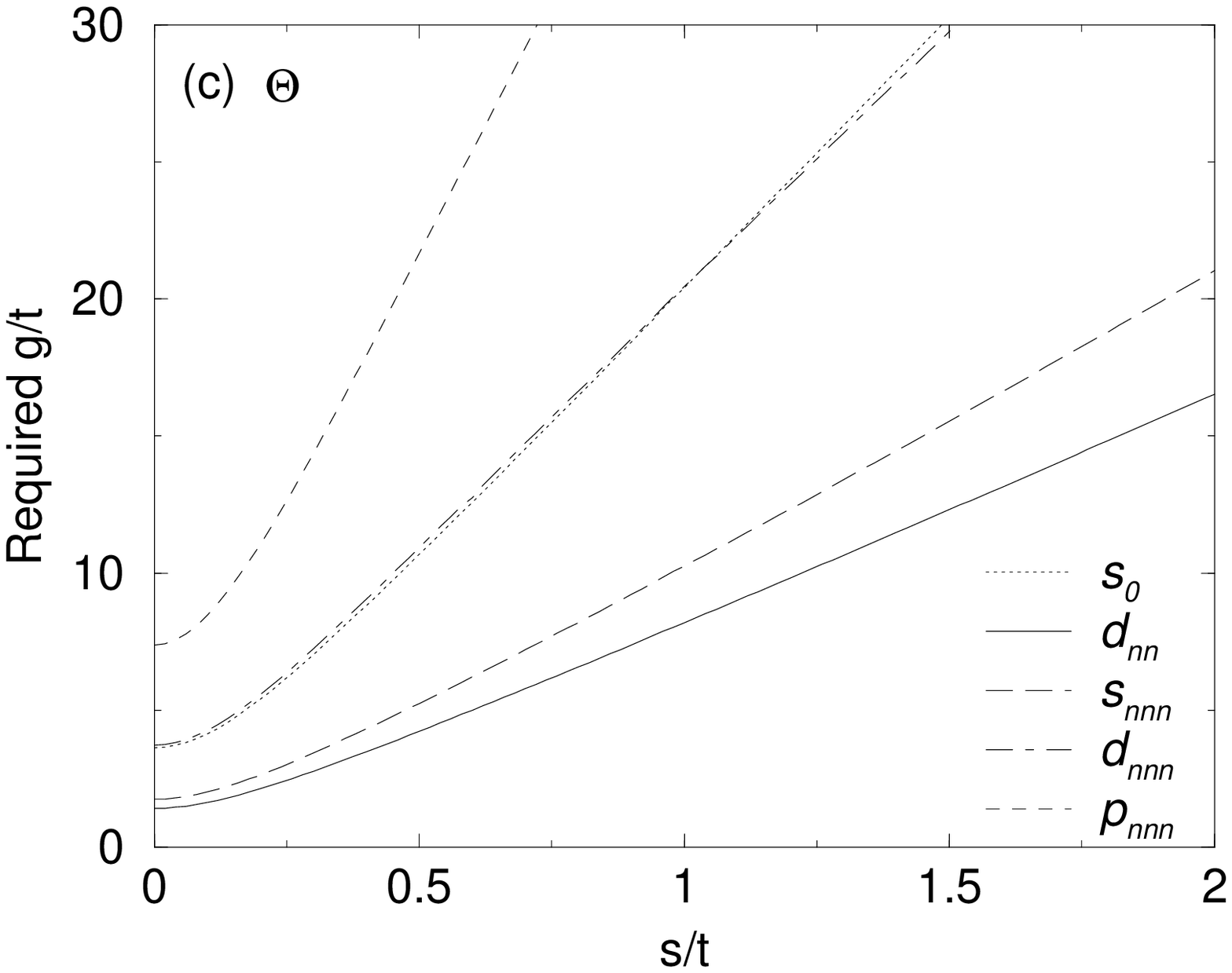}}
\caption{
Comparison of stability of different SC channels in 2D model: required
BCS coupling $g_{\rm BCS}$ to achieve a given superconducting critical
temperature $T_c= 100\mbox{K} (t/1\mbox{eV})$, as a function of the
magnitude of the antiferromagnetic background $s$ at fixed filling $\rho$
for different shapes of the gap, which are listed in
Table~{\protect\ref{etaJ}}.  (For a hopping parameter $t=0.5\,$eV
adequate for HTSC this corresponds to $T_c=50\,$K).  The filling
$\rho=0.9$ considered here is representative.  (a):
$\iDelta\&\iTheta$-channel i.e.\ mixture of staggered and translation
invariant gaps which we find to dominate throughout.  (b):
$\iDelta$-channel i.e.\ pure translation invariant gap.  (c):
$\iTheta$-channel, i.e.\ pure staggered gap.  Note the different scales
of the $g_{\rm BCS}$-axes in the different plots.  There is also an
$s_{\rm nn}$-channel (not shown in the figure) in both the mixed and
purely staggered channel which requires an even higher coupling.
}
\label{fig1}
\end{figure*}

\begin{figure}
\centerline{\epsfxsize=9truecm\epsffile{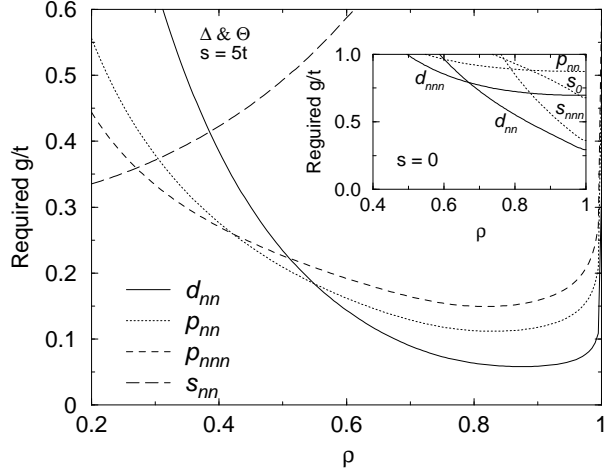}}
\caption{
Required coupling $g_{\rm BCS}$ to reach a critical temperature $T_c=
100\mbox{K} (t/1\mbox{eV})$ as a function of the filling $\rho$ for
fixed magnitude of the AF background, $s=5t$.  The different curves
correspond to different shapes of the gap, for the mixed channels,
$\iDelta \& \iTheta$. The purely translation invariant and staggered
solutions require a much stronger coupling.  The inset shows the
corresponding result for $s=0$ (no AF).  }
\label{fig2}
\end{figure}

\begin{figure}
\centerline{\epsfxsize=9truecm\epsffile{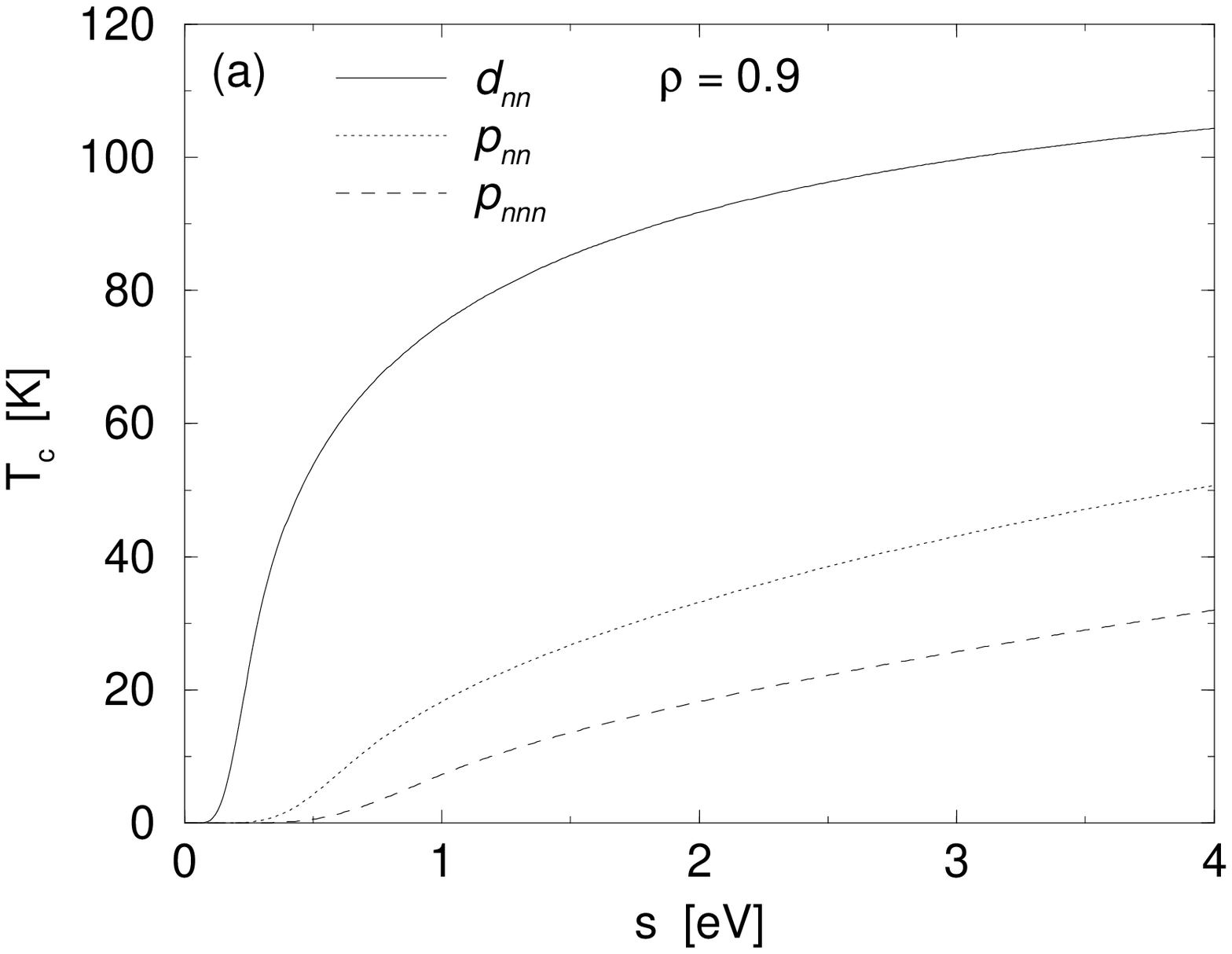}
\epsfxsize=9truecm\epsffile{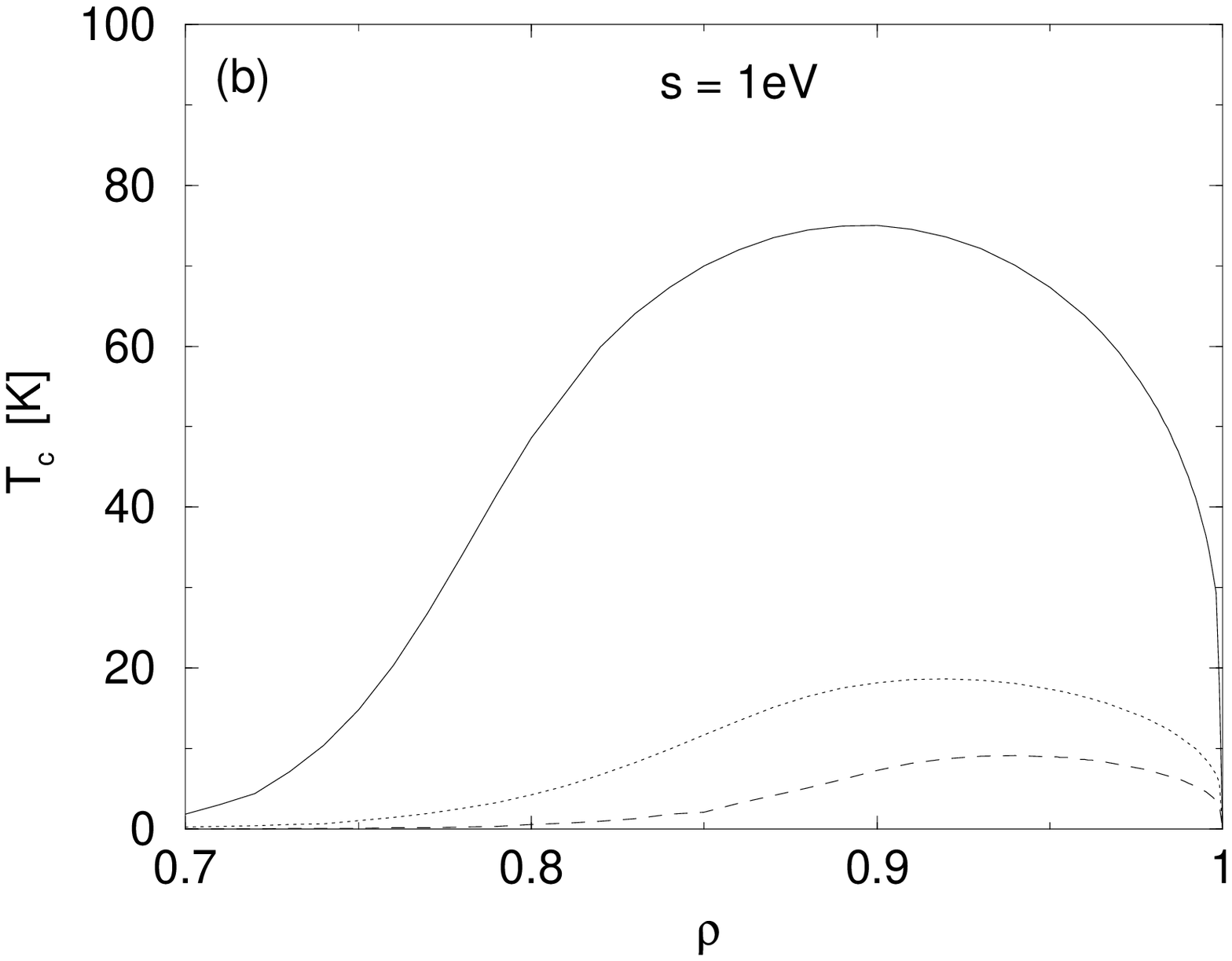}}
\caption{
Critical temperature $T_c$ as a function of magnitude of
antiferromagnetic background $s$ and filling $\rho$, for dominating SC
channel $\iDelta\&\iTheta$ and different shapes of the gap.  Used
parameters here are: $t=0.5\,$eV and $g_{\rm BCS}=0.05\,$eV.  (a)
$\rho=0.9$ fixed (b) $s=1\,$eV fixed.  Labeling of results for
different shapes of the gap as in Fig.~{\protect\ref{fig1}}.  }
\label{fig3}
\end{figure}

The results in Figs.~\ref{fig1}--\ref{fig3} were obtained by
evaluating the $T_c$ equations (\ref{Tc}).  To compare the stability
of different gaps, we found it convenient to determine the required
BCS coupling $\gg$ to achieve a given critical temperature $T_c$.  In
Fig.~\ref{fig1}, the required $\gg$ is plotted as a function of the
magnitude $s$ of the AF background for all possible channels.  We
chose $\rho=0.9$ and $T_c=50\,$K as representative examples.  One can
clearly see that the mixed channels $\iDelta\&\iTheta$ are the only
ones which are enhanced by the antiferromagnetism and always dominate
over the others.  It is also always the same channels which dominate:
$(\iDelta\&\iTheta)^{\rm nn}_{d}$ in case of a nearest neighbor
attraction, and $(\iDelta\&\iTheta)^{\rm nnn}_{p}$ in the next nearest
neighbor case [cf.\ Table \ref{channels}].  The latter one requires a
much larger coupling $\gg$ than the former, though.  These results do
not change qualitatively as the doping is changed over a wide
parameter range.  As $s$ goes to zero the staggered part of the mixed
channels decrease in comparison to the translation invariant part, and
they completely decouple at $s=0$.
This, however, happens only for quite small $s$; for larger $s$ they are of
approximately equal magnitude.

In Fig.~\ref{fig2} the same quantity is plotted as a function of
filling, $\rho$, for fixed $s=5t$.  One can clearly see the optimal
doping away from half filling at which $T_c$ for fixed coupling will
have a maximum.  To illustrate the stabilizing effect of AF
background, the inset in Fig.~\ref{fig2} shows the corresponding
results for $s=0$ (no AF background) for comparison.  As expected, in
this case the most stable SC occurs at half filling where the chemical
potential intersects the van Hove singularity of the 2D DOS, but the
required couplings $\gg$ to reach significant $T_c$ values are much
larger than for nonzero $s$.



We also give some typical examples of $T_c$ as a function of $s$, with
the filling $\rho$ fixed [Fig.~\ref{fig3}(a)] and $T_c$ as a function
of $\rho$, with $s$ fixed [Fig.~\ref{fig3}(b)].  We only show the
dominating mixed channels.  Obviously the antiferromagnetic background can
lead to critical temperatures of the same order of magnitude as
experimental results for HTSC
already for quite small coupling.
Comparing our phase diagram to the experimental results for HTSC,
there are some similarities, e.g., an optimal doping (at $\rho \approx
0.9$ in the nn channel for the parameters considered), but also some
notable differences, e.g., the SC phase in Fig.~\ref{fig3}(b) extends
down to filling $\rho=1$.  However one should keep in mind that the
doping in our model can differ from the exact result if for example
part of the holes do not superconduct due to effects not included in
our treatment.

We have also tested the validity of the BCS-like formulas $T_c=\Omega\,
\exp{(-1/\gg\Lambda)}$.  As discussed, one can expect this formula to be a
good approximation for sufficiently small temperatures $T_c$, but we found
that it works reasonably well up to quite high $T_c$ values (for the
parameters in Fig.~\ref{fig3}(a) up to $T_c\approx 40\,$K).  $\Omega$ in
this formula has the interpretation of an effective energy cut--off for the
interaction, similar to the Debye frequency $\omega_D$ in the simple BCS
model.
We note that for fixed $\rho$, $\Omega$ becomes independent of the shape of
the gap and proportional to the AF band width $\approx t^2/s$ for larger
$s$-values.
For the parameters adequate for high temperature superconductors which we
considered here, $\Omega$ is huge --- of order $1\,$eV$\approx 10^4\,$K ---
for $s=0$, but it becomes comparable to small values typical for standard
superconductors (order $10\,$K) in presence of AF order.  Nevertheless,
without AF order ($s=0$), one never can get a significant $T_c$ since
$\Lambda$ is too small, and the increase of $T_c$ with $s$ is due to the
increase of $\Lambda$ resulting from the increase of the DOS in the AF
bands: Since $\Lambda$ enters $T_c$ exponentially, it overcompensates the
decrease of $\Omega$ with $s$.

\begin{figure}
\centerline{\epsfxsize=9truecm\epsffile{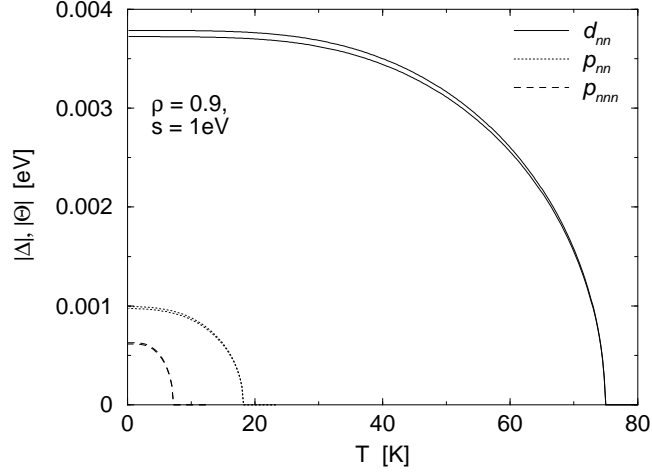}}
\caption{ Magnitude of the superconducting gaps, $|\iDelta|$ and
$|\iTheta|$, as a function of temperature $T$ for dominating channels
$\iDelta\&\iTheta$.  Two neighboring curves show the translation
invariant ($\iDelta$) and staggered ($\iTheta$) parts of the gap.  In
all cases, $|\iDelta|$ is bigger than $|\iTheta|$.  Note that
$|\iDelta|$ and $|\iTheta|$ are nearly the same here, which shows that
the SC gap closely follows the antiferromagnetic background, as
discussed in the text (they become identical for $s\to\infty$).  The
parameters are $\rho=0.9$, $s=1\,$eV, $t=0.5\,$eV and $g=0.05\,$eV.  }
\label{fig4}
\end{figure}

In Fig.~\ref{fig4} we show the temperature dependence of the magnitude of
the superconducting order parameters $\iDelta$ and $\iTheta$ for parameters
values realistic for HTSC [Eqs.~(\ref{MFsol})].  In all cases, $|\iDelta|$
is slightly bigger than $|\iTheta|$.  Note that $|\iDelta|$ and $|\iTheta|$
are nearly the same, which shows that the SC gap closely follows the
antiferromagnetic background.  This is only true for larger $s$--values.
{}From this figure we also obtain $2|\Delta_{\rm max}(T=0)|/k_B
T_c\approx 4.7$ ($\Delta_{\rm max} \equiv \max \Delta({\bf x},{\bf y})
= 2(|\iDelta|+|\iTheta|)$), but we found that this parameter depends
on doping and $s$ and is not universal as in the BCS model.

\begin{figure}
\centerline{\epsfxsize=9truecm\epsffile{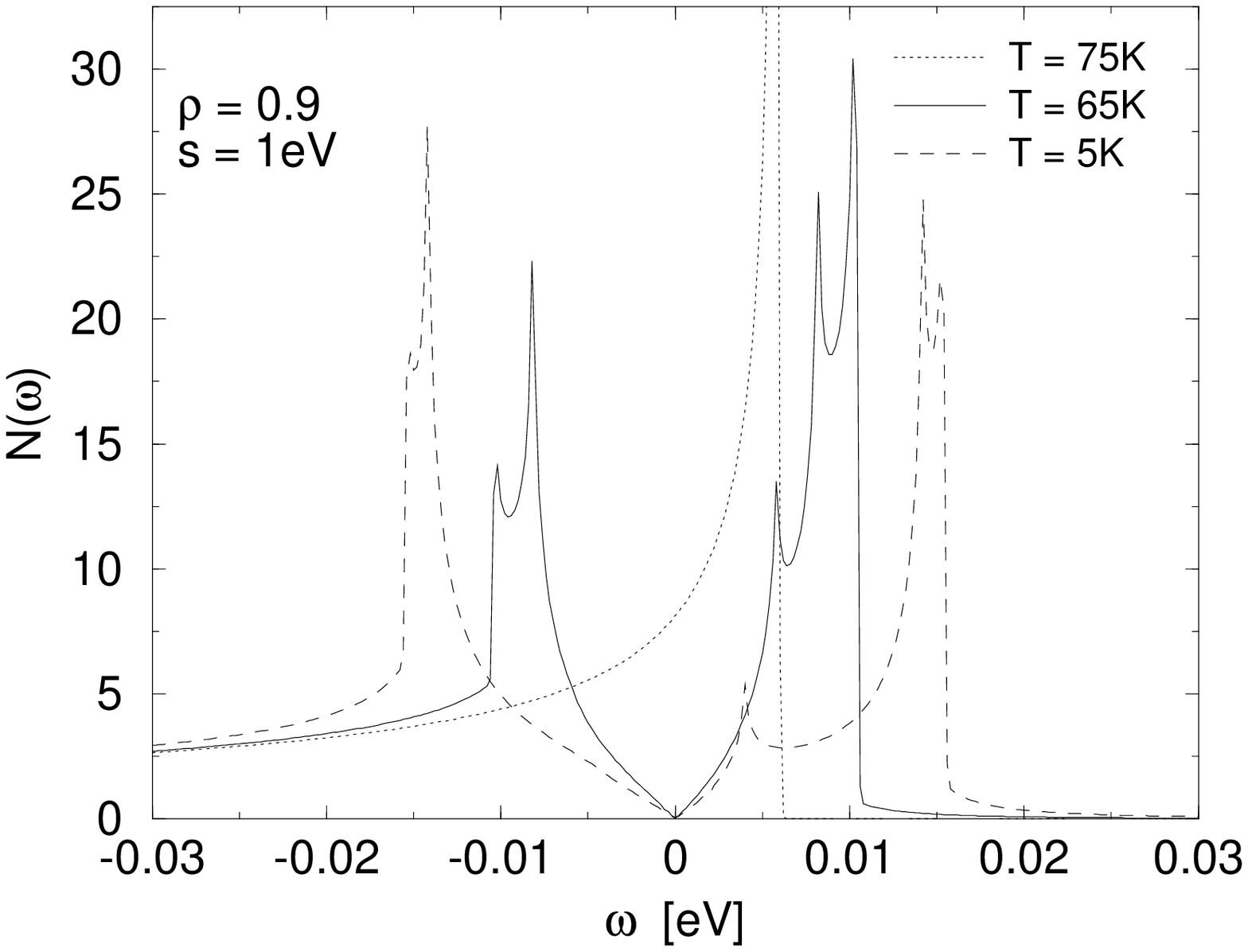}
\epsfxsize=9truecm\epsffile{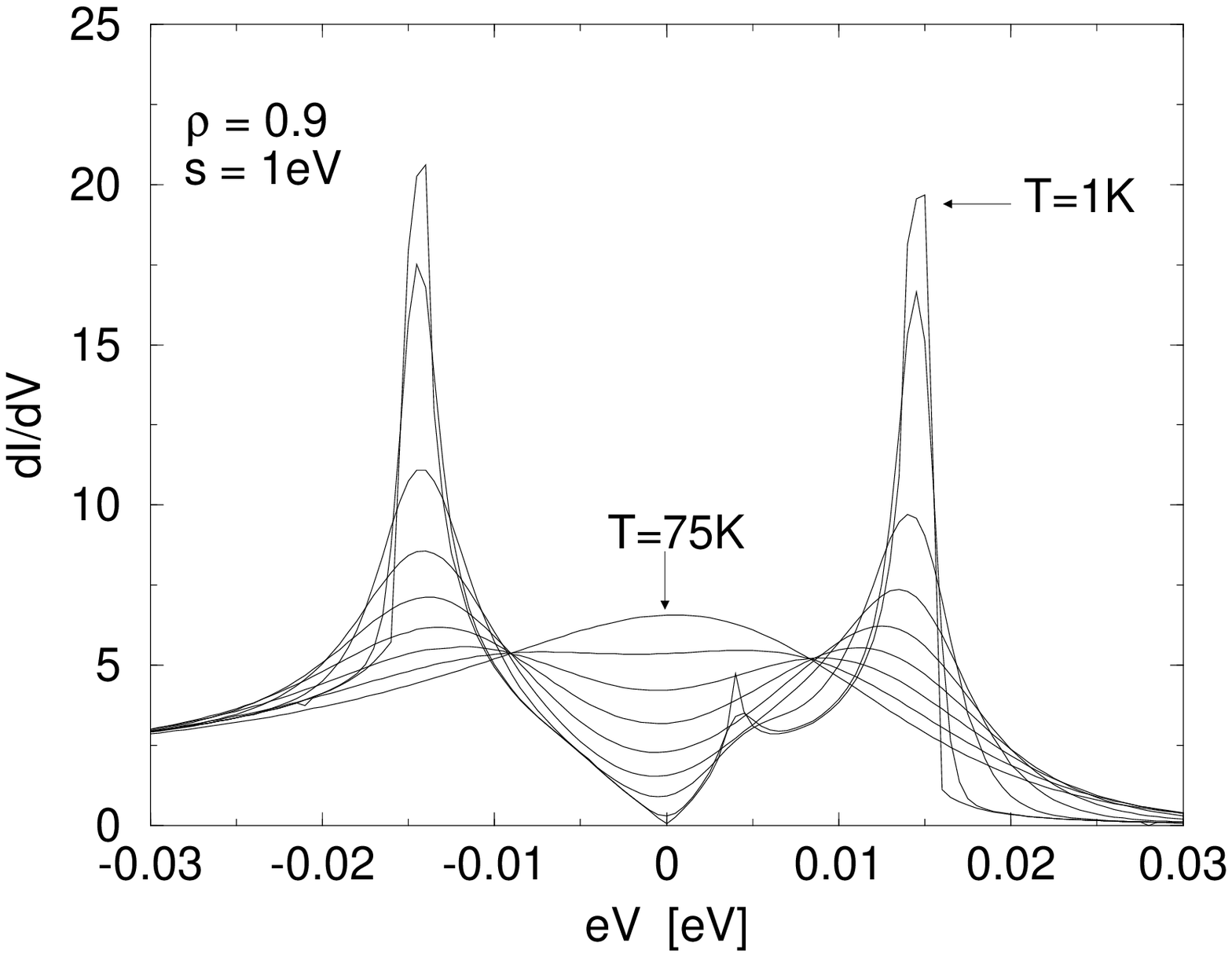}}
\caption{ (a) Superconducting density of states for three different
temperatures $T = 75\,$K, $65\,$K and $1\,$K.  The parameters are the
same as in Fig.~{\protect\ref{fig4}}, with a $T_c \approx 75\,$K.  The
chemical potential is at $\omega = 0$.  (b) Tunneling conductance
$dI/dV$ for a superconductor -- normal-metal junction as a function of
the applied voltage for a sequence of temperatures $T = 75, 65, 55
\ldots, 5\,$K and $1\,$K.  Note that most of the fine structures in
the density of states is washed out by temperature effects.}
\label{tunnel}
\end{figure}

Another interesting quantity is the superconducting one-particle
electronic density of states $N(\omega)$.  In Fig.~\ref{tunnel}(a) we
show how $N(\omega)$ changes as the temperature is lowered from $T_c$,
for a $d_{\rm nn}$-wave gap [Eq.~(\ref{NE})].  Here one sees how the SC
gap opens up at the Fermi level, which is at $\omega=0$.  The filling
used in the figure is chosen to give an optimal $T_c$ for this value
of $s$, and one clearly sees that the gap opens up precisely at the AF
peak in $N(\omega)$ as the temperature is lowered below $T_c$.  It is
interesting that our simple model can lead to the rich structure in
the DOS seen in the figure.  In Fig.~\ref{tunnel}(b) we show the
tunneling conductance, $dI/dV$, for a superconductor -- normal-metal
junction, which is essentially the same curve smeared out by positive
temperature effects [Eq.~(\ref{dIdV})].  Here we note some significant
features also seen in experiments \cite{tunn}: (1) There is not a real
gap but only a dip in the tunnel conductance.  This is a consequence
of the $d$-wave shape of the gap, which implies that it has nodes on
the Fermi surface.  (2) The positions of the two peaks do not move
much as the temperature is lowered, even though they would be expected
to, because of the increase of the gap.  It is clear, especially when
compared to Fig.~\ref{tunnel}(a), that this comes from thermal
smearing.  At low temperature, when the smearing is small, the gap is
essentially constant as a function of temperature.  Some details of
the structure in the DOS remain visible at very low temperature.  Note
also that the separation of the two peaks is about $2|\Delta_{\rm max}|$
(compare Figs.~\ref{fig4} and \ref{tunnel}).

The results also provide a consistency check for our assumption
that the different SC channels do not mix below $T_c$:
for larger $s$ values, it is always the channel
$\iDelta\&\iTheta^{\rm nn}_{d}$ respectively $\iDelta\&\iTheta^{\rm
nnn}_{p}$ which is dominating: for a given coupling $g_{\rm nn}$
respectively $g_{\rm nnn}$, it always leads to a $T_c$ much bigger
than the others, and also the gaps for the different channels remain
well-separated for all temperatures. This suggests that
neglecting the channel mixing was justified.

\section{Discussion} \label{conclusion}

In this paper we studied a simple model of superconductivity in an
antiferromagnetic background and we analyzed in detail the two
dimensional version of this model.  Our model offers a simple account
for one effect of AF correlations on superconductivity: AF can narrow
bands and thereby boost $T_c$.  We also discussed the possible
relevance of this for HTSC, although the connection is not firmly
established.

To summarize our results, in case of nearest neighbor (nn) attraction,
we found that the internal structure of the gap is $d$-wave, whereas
for next-nearest neighbor (nnn) attraction it is $p$-wave.  As soon as
AF is present, the same coupling produces a much larger $T_c$ in the
$d$-wave than in the $p$-wave case, and on-site attraction leads only
to insignificant $T_c$ values.  The most stable SC channels correspond
to paired fermions oriented parallel to the AF order parameter, and in
case the paired fermions are on adjacent sites (nn) they have opposite
spins and correspond to a mixture of spin singlet and triplet.  The
latter is possible because spin rotation symmetry is broken.  In the
$d$-wave channel, we found that for reasonably weak attraction,
switching on the AF order increased $T_c$ to the right order of
magnitude for HTSC.

We stress that all the results presented in this paper are within a
mean field approximation. We expect fluctuation corrections to the
curves in all the figures, and the results should thus be regarded
only as order of magnitude estimates demonstrating the qualitative
features of the model.

A comment about the apparent long range order of our mean field
solution is in order.  The mean field analysis done here allows to
determine only the {\em magnitude} of the SC order parameter but {\em
not} its {\em phase}, and our calculated $T_c$ rather corresponds to
the mean-field transition temperature $T_{c0}$.  Thus nontrivial
solutions of our mean field equations do not contradict the
nonexistence of continuous symmetry breaking in two dimensions.  A
complete analysis of SC in two dimensions also has to take into
account phase fluctuations \cite{EK} and is beyond the scope of this
paper.

To conclude, we presented a simple model for electrons in an AF
background, and studied superconductivity in this model.  We discussed
how this model may describe certain aspects of the physics of
antiferromagnetically correlated fermions, as present e.g.\ in HTSC.
Some of our results are qualitatively similar to certain basic
properties of HTSC.  The methods presented here should be useful also
for other similar and more complicated models, in particular for cases
where several effective bands contribute to superconductivity.

\acknowledgments
We thank Asle Sudb{\o} for discussions.  This work was supported by
the Swedish Natural Science Research Council.

\appendix

\section{On  Staggered States}
\label{AppA}
In this Appendix we show how to evaluate the inverse for Green
functions etc.\ describing staggered states.  Note that the quantities
$A(x,y)$ etc.\ below are matrices and the matrix product is
understood.

To get the inverse of $A(x,y)= A_N(x-y) + A_A(x-y)e^{{i} Qy}$, we
solve the equation
\eqa
\sum\limits_z \bigl( A_N(x-z) + A_A(x-z)e^{{i} Qz}\bigr)\nonu
\times \bigl( B_N(z-y) + B_A(z-y)e^{{i} Qy}\bigr) = \delta_{x,y} \; ,
\eqaend
which, after Fourier transformation, can be written in a matrix form
\eq
\left( \matrix{A_N & A_A \cr A_A^Q & A_N^Q } \right)
\left( \matrix{B_N & B_A \cr B_A^Q & B_N^Q } \right) =
\left( \matrix{ \bf 1 & \bf 0 \cr \bf 0 & \bf 1 } \right)
\eqend
where $A_N^Q(k) = A_N(k-Q)$, etc. The solutions are (suppressing the
argument $k$)
\eqa
B_N &=& ( A_N - A_A (A_N^Q)^{-1} A_A^Q )^{-1} \nonu
B_A &=& ( A_A^Q - A_N^Q A_A^{-1}A_N)^{-1}
\eqaend
If all of $A_N,A_A,B_N,B_A$ commute,
\eqa
B_N &=& \frac{A_N^Q}{A_N A_N^Q - A_A A_A^Q} \nonu
B_A &=& \frac{-A_A}{A_N A_N^Q - A_A A_A^Q}
\eqaend
In particular, this gives the free Green function (\ref{Gn}) as the inverse
of $R = G_0^{-1}$ has the translation invariant and staggered parts
$R_{N,A}$ given in Eq.\ (\ref{QNA}).

\section{$T_c$ equations}
\label{AppB}
Here we give some details of our derivation of the $T_c$ equation
in Section \ref{Tceqs}.

We use the following notation for functions of $k$: \[f^Q(k) = f(k -
Q), \quad \tilde f(k) = - f(-k)\] We also write $G_N(k) =
g(k)\sigma_0,\ G_A(k) = a(k) \sigma_3$ and use that
$\eta^Q(k)=\eta(k-Q)= p\,\eta(k)$, where $p =\pm 1$ depending on the
parity of $\iDelta(k)$.

With that we obtain the equations
\neqa
\iDelta_{1,4} = -g_{\rm BCS}\sum_k\Bigl( \bigl( {g\tilde g}\eta^2
+ {a\tilde a^Q}\eta\eta^Q
\bigr)\iDelta_{1,4}\\
\pm \bigl(  {g\tilde a^Q}\eta^2 + {a\tilde g}\eta\eta^Q
\bigr)\iTheta_{1,4}\Bigr) \\
\iTheta_{1,4} = -g_{\rm BCS}\sum_k\Bigl(\bigl( {g\tilde g^Q}\eta^2 + {a\tilde
a}\eta\eta^Q
\bigr)\iTheta_{1,4}\\
\pm \bigl(  {g\tilde a}\eta\eta + {a\tilde g^Q}\eta\eta^Q
\bigr)\iDelta_{1,4}\Bigr)
\neqaend
and
\neqa
\iDelta_{2,3} = -g_{\rm BCS}\sum_k\Bigl( \bigl({g\tilde g}\eta^2 - {a\tilde
a^Q}\eta\eta^Q
\bigr)\iDelta_{2,3}\\
\mp \bigl(  {g\tilde a^Q}\eta^2 - {a\tilde g}\eta\eta^Q
\bigr)\iTheta_{2,3}\Bigr) \\
\iTheta_{2,3} = -g_{\rm BCS}\sum_k\Bigl( \bigl( {g\tilde g^Q}\eta^2 - {a\tilde
a}\eta\eta^Q
\bigr)\iTheta_{2,3}\\
\mp \bigl(  {g\tilde a}\eta^2 - {a\tilde g^Q}\eta\eta^Q
\bigr)\iDelta_{2,3}\Bigr)
\neqaend
Doing the Matsubara sums and defining
\eq
\chi_\beta(E) \equiv \f{1}{\beta}\sum_{\omega_n}\f{1}{\omega_n^2 + E^2} =
\f{\tanh(\beta E/2)}{2E}
\eqend
results in Eq.\ (\ref{Tc}) with the constants
\eqa
\label{betarep}
K_+(\beta) &=&
\intk\Bigl( \f{1}{2} \bigl(\chi_\beta(E_+) + \chi_\beta(E_-)\bigr)
\eta_J^2\Bigr) \nonu
K_-(\beta) &=&
\intk\Bigl( \f{1}{2} \bigl(\chi_\beta(E_+) + \chi_\beta(E_-)\bigr)
\eta_J^2 \nonu
&&-\f{s^2}{2\mu w}\bigl(\chi_\beta(E_+) -  \chi_\beta(E_-)\bigr) \eta_J^2
\Bigr) \nonu
\tilde K_+(\beta)&=&
\intk\Bigl( \f{1}{2} \bigl(\chi_\beta(E_+) + \chi_\beta(E_-)\bigr) \eta_J^2
\nonu
&&-\f{\epsilon^2}{2\mu w}\bigl(\chi_\beta(E_+) -  \chi_\beta(E_-)\bigr) \eta_J^2
\Bigr) \nonu
\tilde K_-(\beta)&=&
\intk\Bigl( \f{1}{2} \bigl(\chi_\beta(E_+) + \chi_\beta(E_-)\bigr) \eta_J^2
\nonu
&&-\f{w}{2\mu}\bigl(\chi_\beta(E_+) -  \chi_\beta(E_-)\bigr) \eta_J^2
\Bigr) \nonu
L_+(\beta) &=&
\intk\Bigl( \f{s}{2w}\bigl(\chi_\beta(E_+) - \chi_\beta(E_-)
\bigr) \eta_J^2\Bigr)  \nonu
L_-(\beta) &=& 0
\eqaend
where $w =\sqrt{\epsilon^2+s^2}$ and $E_\pm = -\mu\pm w$. Note that
all these expressions have well-defined limits $\mu\to 0$.

It is shown below (and can be checked numerically) that the $\beta$
dependence of these integrals for large $\beta$ is in a very good
approximation given by (these formulas becomes exact for
$\beta\to\infty$ )
\eqa
\label{betadep}
K_+(\beta) =M\log(\Omega^{(a)}\beta),&&\quad  K_-(\beta) = \tilde\alpha \nonu
\tilde K_+(\beta) = \f{s^2}{\mu^2} M\log(\Omega^{(b)}\beta)
+\alpha,&&\quad \tilde K_-(\beta) = \alpha+\tilde \alpha \\
L_+(\beta) = \f{s}{\mu }M\log(\Omega^{(c)}\beta) \nonumber
\eqaend
with $M$ Eq.\ (\ref{M}) and $\alpha=\alpha_\beta$ and
$\tilde\alpha=\tilde\alpha_\beta$ are, in a very good approximation,
independent of $\beta$ for large $\beta$ [the latter can be checked
numerically; the energy scales $\Omega^{(a,b,c)}$ can be obtained by
calculating the integrals $K_+(\beta)$ etc.\ for one (large)
$\beta$-value numerically and comparing with (\ref{betadep}).]  This
yields the $T_c$-equations of the form Eq.\ (\ref{bcs}).

The step going from (\ref{betarep}) to (\ref{betadep}) uses that
functions $F(E)$ reasonably smooth close to $E=0$, the
$\beta$-dependence of
\eq
I(F,\beta)= \int{d} E\, F(E) \f{\tanh(\beta E/2)}{2E}
\eqend
for large $\beta$ is approximated well by
\eq
\label{LogEq}
I(F,\beta) = F(0)\log\bigl(\Omega_F \beta \bigr)
\eqend
To see this, define
\[
\Omega(F,\beta) = \frac{1}{\beta} e^{I(F,\beta)/F(0)},
\]
then $I(F,\beta) = F(0)\log\bigl(\Omega(F,\beta) \beta \bigr) $.  Now,
with $x=\beta E/2$,
\neqa
\f{d\Omega}{d\beta} = \frac{\Omega}{\beta } \Bigl( \int dx
\frac{F(2x/\beta)}{F(0)}\frac{1}{2 \cosh^2 (x)} -1 \Bigr)
\neqaend
Taylor expanding $F$, we see that $d\Omega/d\beta=O(1/\beta^3)$ so we
can replace $\Omega(F,\beta)$ by $\Omega_F = \lim\limits_{\beta \to
\infty} \Omega(F,\beta)$ up to terms of order $1/\beta^2$.

\end{document}